%% file: HPR_OS.tex
\documentclass[11pt,a4paper,twoside,fleqn]{article}
\usepackage{color,graphicx,lscape,here,geometry,setspace,multirow,enumitem}
\usepackage[dvipsnames]{xcolor}
\usepackage{graphicx}
\usepackage{natbib}
\usepackage[utf8, latin1]{inputenc}
\usepackage{authblk}
\usepackage{silence}
\usepackage{tgpagella}
\usepackage{helvet}

\usepackage{rotating}
\usepackage{multirow}

\usepackage{booktabs}
\usepackage{diagbox}
\usepackage{color,colortbl}
\usepackage{setspace}
\usepackage{algorithm2e}
\usepackage{enumitem}
\usepackage{amsfonts}
\usepackage{tikz}

\definecolor{nblue}{HTML}{000660}
\usepackage[colorlinks=true,urlcolor=nblue,linkcolor=nblue,citecolor=nblue]{hyperref}
\usepackage{bigstrut}

\usepackage{threeparttable}
\usepackage{dcolumn}
\newcolumntype{d}[1]{D{.}{.}{#1}}

\usepackage{array}
\newcolumntype{C}[1]{>{\centering\arraybackslash}p{#1}}

\usepackage{etoolbox} 
\makeatletter
\patchcmd{\BR@backref}{\newblock}{\newblock[}{}{}
\patchcmd{\BR@backref}{\par}{]\par}{}{}
\makeatother

\usepackage{pdflscape}
\usepackage{afterpage}

\usepackage{longtable}
\usepackage{array}

\usepackage{comment}
\usepackage{mathtools}
\usepackage[hang,flushmargin]{footmisc} 
\usepackage[hang]{footmisc}
\usepackage[british]{babel}

\pdfoutput=1
\usepackage{float}
\restylefloat{table}

\usepackage[title,titletoc]{appendix}
\makeatletter

\renewenvironment{appendices}{%
    \begin{oldappendices}%
    \renewcommand{\thefigure}{\ifnum \c@section>\z@ \thesection.\fi\@arabic\c@figure}%
    \@addtoreset{figure}{section}%
    \renewcommand{\thetable}{\ifnum \c@section>\z@ \thesection.\fi\@arabic\c@table}%
    \@addtoreset{table}{section}}{%
    \end{oldappendices}%
}\makeatother

\usepackage{titlesec} 
\titleformat{\section}[block]{\large\bfseries}{\thesection. }{0em}{\MakeUppercase} 
\titleformat{\subsection}[block]{\large}{\thesubsection. }{0em}{\itshape} 
\titleformat{\subsubsection}[block]{\large}{}{0em}{\itshape} 

\let\natbibcitet\citet
\renewcommand\citet{\bibpunct{(}{)}{,}{a}{,}{,}\natbibcitet}
\let\natbibcitep\citep
\renewcommand\citep{\bibpunct{(}{)}{;}{a}{,}{;}\natbibcitep}
\newcommand{\bi}{\begin{itemize}}
\newcommand{\ei}{\end{itemize}}
\newcommand{\be}{\begin{equation}}
\newcommand{\ee}{\end{equation}}
\defcitealias{ieo14}{IEO, 2014}

\long\def\symbolfootnote[#1]#2{\begingroup%
\def\thefootnote{\fnsymbol{footnote}}\footnote[#1]{#2}\endgroup}

\makeatletter
\def\ubar#1{\underline{\sbox\tw@{$#1$}\dp\tw@\z@\box\tw@}}
\def\obar#1{\overline{\sbox\tw@{$#1$}\dp\tw@\z@\box\tw@}}
\makeatother

\usepackage{bm}
\usepackage{caption}
\usepackage{subcaption}

\makeatletter\let\p@subfigure\thefigure\makeatother

\floatstyle{plaintop}
\restylefloat{table}

\usepackage{fancyhdr} 
\usepackage{cleveref}
\crefname{chapter}{Chapter}{Chapters}
\crefname{section}{Section}{Sections}
\crefname{subsection}{Section}{Sections}
\crefname{subsubsection}{Section}{Sections}
\crefname{figure}{Figure}{Figures}
\crefname{table}{Table}{Tables}
\crefname{equation}{Equation}{Equations}
\crefname{appendix}{Appendix}{Appendices}
\crefname{appendices}{Appendix}{Appendices}
\crefname{appsec}{Appendix}{Appendices}

\makeatletter

\def\Autoref#1{%
  \begingroup
  \edef\reserved@a{\cpttrimspaces{#1}}%
  \ifcsndefTF{r@#1}{%
    \xaftercsname{\expandafter\testreftype\@fourthoffive}
      {r@\reserved@a}.\\{#1}%
  }{%
    \ref{#1}%
  }%
  \endgroup
}
\def\testreftype#1.#2\\#3{%
  \ifcsndefTF{#1autorefname}{%
    \def\reserved@a##1##2\@nil{%
      \uppercase{\def\ref@name{##1}}%
      \csn@edef{#1autorefname}{\ref@name##2}%
      \autoref{#3}%
    }%
    \reserved@a#1\@nil
  }{%
    \autoref{#3}%
  }%
}
\makeatother

\makeatletter
\newlength{\qrr@dimen@}
\expandafter\pretocmd\csname tabular*\endcsname{\setlength{\qrr@dimen@}{#1}}{}{}
\newcommand*{\Rowcolor}[2][\tabcolsep]{%
    \ifx\relax#1\relax\else
        \kern-\the\dimexpr#1\relax
    \fi
    \makebox[0pt][l]{%
        \fboxsep=0pt
        \colorbox{#2}{%
            \strut\kern\qrr@dimen@
        }%
    }%
    \ifx\relax#1\relax\else
        \kern\the\dimexpr#1\relax
    \fi
    \ignorespaces
}
\makeatother

\definecolor{red}{rgb}{0.95, 0.3, 0.3}
\definecolor{grey}{rgb}{0.9, 0.9, 0.9}

\providecommand{\shadeRow}{\Rowcolor[0pt]{grey}}
\providecommand{\shadeBench}{\Rowcolor[0pt]{red}}

\def\titlestring{Sparse time-varying parameter VECMs with an application to modeling electricity prices}
\title{\sffamily\LARGE{\textbf{\titlestring}}}

\author[1]{Niko \MakeUppercase{Hauzenberger}}
\author[2]{Michael \MakeUppercase{Pfarrhofer}\footnote{\textit{Correspondence to}: Michael Pfarrhofer, email: \href{mailto:mpfarrho@gmail.com}{mpfarrho@gmail.com}. We thank Florian Huber, Matteo Iacopini, Michael McCracken and Francesco Ravazzolo for valuable comments and suggestions. Hauzenberger and Pfarrhofer acknowledge financial support from the Austrian Science Fund (FWF, grant no. \texttt{ZK 35}) and the Oesterreichische Nationalbank (OeNB, Jubil\"aumsfonds, projects \texttt{18127}, \texttt{18763} and \texttt{18765}).}}
\author[3]{Luca \MakeUppercase{Rossini}}
\affil[1]{\textit{University of Salzburg and University of Strathclyde}}
\affil[2]{\textit{University of Vienna and WU Vienna}}
\affil[3]{\textit{University of Milan}}
\date{}


\def\equationautorefname~#1\null{%
  Eq.~(#1)\null
}

\begin{document}
\maketitle\onehalfspacing\thispagestyle{empty}\normalsize\vspace*{-2em}\small

\begin{center}
\begin{minipage}{0.8\textwidth}
{\noindent\small In this paper we propose a time-varying parameter (TVP) vector error correction model (VECM) with heteroskedastic disturbances. We propose tools to carry out dynamic model specification in an automatic fashion. This involves using global-local priors, and postprocessing the parameters to achieve truly sparse solutions. Depending on the respective set of coefficients, we achieve this via minimizing auxiliary loss functions. Our two-step approach limits overfitting and reduces parameter estimation uncertainty. We apply this framework to modeling European electricity prices. When considering daily electricity prices for different markets jointly, our model highlights the importance of explicitly addressing cointegration and nonlinearities. In a forecast exercise focusing on hourly prices for Germany, our approach yields competitive metrics of predictive accuracy.}
\end{minipage}
\end{center}
\normalsize\renewcommand{\thepage}{\arabic{page}}\newpage\doublespacing

\section{Introduction}\label{sec:intro}
This paper discusses econometric tools to achieve dynamic model specification for vector error correction models (VECMs) automatically. Our main idea is to start with a suitably flexible and sophisticated specification amongst this model class, and to impose data-driven shrinkage on the parameter space to obtain the simplest adequate nested version. This approach is motivated by our applications, where no clear theoretical guidance is available about how to choose crucial modeling aspects deterministically. Using a very general model guards against underfitting and misspecification, while pushing its parameters towards a simpler sparsified solution avoids overfitting and poor out-of-sample (OOS) predictive performance.

Specifically, we propose a time-varying parameter (TVP) VECM with heteroskedastic errors and apply it to model and forecast European electricity prices. Deregulation and increasingly competitive markets in the power sector have led to a surge of interest in statistical methods for modeling and forecasting electricity demand and price dynamics. Competing approaches include both univariate and multivariate time series models in linear and nonlinear settings (e.g., structural breaks in the conditional means and variances). For an overview of the related literature, see \citet{WERON20141030}. Most directly related to our approach, \citet{DeVany1999} show that electricity prices in US states are cointegrated, with long-run relationships driven by no-arbitrage conditions. The more recent literature also finds evidence in favor of common dynamics and cointegration between electricity and gas prices for major power exchanges in the European Union \citep[see, among others,][]{Bosco2010,Houllier2012,Bello2013,Marcos2016,gianfreda2019Energy}.

We aim at capturing these empirical features and regularities in energy markets and model them explicitly. This motivates our proposed TVP-VECM. However, estimating VECMs, particularly with TVPs, poses several econometric challenges \citep[see, e.g.,][]{koop2011bayesian}. First, even when it is agreed upon that cointegration should be taken into account, there is often no compelling argument about how a set of (economic) variables are cointegrated (especially in higher dimensions). This complicates introducing reasonable restrictions to identify the long-run behavior of such time series. For electricity prices, this aspect is even more apparent, since there is no clear intuition from economic theory about how to \textit{a priori} restrict the cointegration space. Second, and relatedly, the cointegration rank is unknown and may be subject to change over time depending on the application. The previous literature often conditions on the rank and then compares measures of model fit \textit{ex post} \citep[see][]{geweke1996bayesian}. This may be impractical, either due to computational limits, or due to varying the rank requiring additional (possibly \textit{ad hoc}) identifying restrictions.\footnote{Notable exceptions are \citet{jochmann2015regime} and \citet{chua2018bayesian}, who use regime-switching models to estimate a time-varying cointegration rank.} Third, due to their flexibility, large TVP models are prone to overfitting. Many papers thus propose to restrict the parameter space relying on hierarchical prior distributions or approximations.\footnote{These three aspects are discussed to a varying extent in, e.g., \citet{bunea2012joint,jochmann2013stochastic,eisenstat2016stochastic,huber2019threshold,chakraborty2016bayesian,huber2020bayesian, hauzenberger2020dynamic, huber2018stochastic}.}

Our approach to solving these interrelated issues combines several recent econometric techniques used for large-scale TVP models and reduced rank regressions. We employ continuous global-local priors for pushing the parameter space towards sparsity. \citet{pruser2023data} is a related paper with a similar approach to deal with shrinkage in (constant parameter) VECMs. However, as noted by \citet{chakraborty2016bayesian}, such priors solely achieve approximate zeroes; the probability of observing exact zeroes is zero. In simple terms, this implies that shrinkage may guide a general model specification towards a simpler one, but it cannot achieve the nested version exactly (in terms of model selection). 

As a remedy, we postprocess our posterior via minimizing distinct least absolute shrinkage and selection operator (\textsc{Lasso})-type loss functions to obtain truly sparse estimates that may feature exact zeroes. The choices of loss functions are due to different implications of varying sparsity patterns across partitions of the parameter space. In particular, we propose to use distinct loss functions for the cointegration matrix (grouped \textsc{Lasso}), the autoregressive parameters (element-wise \textsc{Lasso}), and the covariances (graphical \textsc{Lasso}). A key feature of our framework is that it selects the number of cointegration relationships for each period, limiting the need for imposing \textit{ad hoc} restrictions \textit{a priori}. Moreover, sparsifying the coefficients \textit{ex post} alleviates overfitting concerns and reduces parameter estimation uncertainty \citep[see also][]{huber2020inducing}.

We apply our model framework in two different energy-related contexts. The first uses daily data from several European markets jointly. Our approach detects distinct patterns in both dynamic and static interdependencies across markets (i.e., the procedure selects relevant relationships in a multicountry context, see, e.g., \citealp{feldkircher2022approximate}), and the results corroborate previous empirical evidence about the importance of addressing cointegration, nonlinearities and heteroskedasticity. In our second application, we conduct an extensive OOS forecast exercise for hourly electricity prices for Germany. In this case, our approach uncovers and/or excludes intra-day relationships between energy prices within a single country. We consider forecasts for each hour of the following day, and find that multivariate cointegration models with TVPs and heteroskedastic errors provide improvements relative to various simpler benchmarks. In particular, the forecast exercise indicates that our proposed sparsified TVP-VECM yields competitive and in many cases superior forecasts for German hourly electricity prices.

Summarizing, the VECM allows for discriminating between long-run equilibria and short-run adjustment dynamics, which we find to be important for modeling electricity prices. TVPs capture structural breaks in the dynamic relationships of the underlying data. This is particularly useful when addressing complex latent pricing mechanisms and varying importance of variables such as fuel prices that may be subject to change over time. Moreover, our \textit{shrink-then-sparsify} approach allows for specification search and variable selection in high-dimensional data by imposing exact zeroes in the coefficient matrices. Heteroskedastic errors capture large unanticipated shocks in prices, a crucial feature when interest centers on producing accurate density forecasts \citep[see, e.g.,][]{gianfreda2020large}.

The paper proceeds as follows. Section \ref{sec:econometrics} presents a flexible TVP-VECM model equipped with global-local shrinkage priors and heteroskedastic errors, while Section \ref{sec:sparsification} discusses our proposed dynamic sparsification techniques. Section \ref{sec:application} applies the TVP-VECM to modeling European electricity prices. Section \ref{sec:conclusion} summarizes and concludes the paper.

\section{Time-varying parameter vector error correction models}\label{sec:econometrics}
We begin by introducing our baseline econometric framework using rather general notation. This reflects the notion that while these methods are developed in light of our applications, they are also applicable in other contexts. Let $\bm{y}_t$ be an $M\times1$ vector of endogenous variables for $t=1,\hdots,T$, and denote the first difference operator by $\Delta$ such that $\Delta \bm{x}_t = \bm{x}_t - \bm{x}_{t-1}$. A general specification of the TVP-VECM is:
\begin{equation}
	\Delta \bm{y}_t = \bm{\Pi}_{t} \bm{w}_{t}+\sum_{p=1}^P \bm{A}_{pt}  \Delta \bm{y}_{t-p} + \bm{\gamma}_t \bm{c}_t + \bm{\epsilon}_t, \quad \bm{\epsilon}_t \sim \mathcal{N}(\bm{0}, \bm{\Sigma}_t).\label{eq:tvpvecm}
\end{equation}
Here, $\bm{w}_t = (\bm{y}_{t-1}',\bm{f}_{t-1}')'$, where $\bm{y}_{t-1}$ is the first lag of the endogenous variables and $\bm{f}_{t-1}$ are a set of $q_f$ exogenous factors such that $\bm{w}_t$ is of size $q\times1$ with $q=M+q_f$. 

The left-hand side of Eq. (\ref{eq:tvpvecm}) is stationary, i.e., integrated of order zero or $I(0)$; this in turn requires the product $\bm{\Pi}_{t}\bm{w}_t$ to be $I(0)$. Assuming unit roots of the endogenous variables in levels, this implies that $\bm{\Pi}_{t}$ is an $M\times q$ matrix of reduced rank. The rank ${r}_t$ reflects the number of linearly independent cointegrating relationships, with ${r}_t<M$. $\bm{A}_{pt}$ refers to an $M\times M$ time-varying coefficient matrix related to the $p$th lag $\Delta \bm{y}_{t-p}$, and $\bm{\gamma}_t$ of size $M\times N$ relates an $N\times1$ vector $\bm{c}_t$ of deterministic terms (such as seasonal dummies, trends, or intercepts) to $\Delta \bm{y}_t$. In our baseline version of the model, we assume a zero mean Gaussian error term $\bm{\epsilon}_t$ with an $M\times M$ time-varying covariance matrix $\bm{\Sigma}_t$.

\subsection{The cointegration matrix}
A more thorough discussion of the matrix $\bm{\Pi}_{t}$ of reduced rank ${r}_t$, governing the cointegration relationships, is in order. In most applications using VECMs, ${r}_t = \bar{r}$ is some fixed (time-invariant) integer with $1\leq\bar{r}\leq(M-1)$. The rank order is commonly motivated either based on economic theory \citep[see, e.g.,][]{giannone2019priors}, determined by calculating marginal likelihoods for a set of possible choices \citep[see, e.g.,][]{geweke1996bayesian}. For large-scale models, these approaches are often computationally prohibitive and rather restrictive.\footnote{Notable exceptions are \cite{huber2019threshold} and \cite{pruser2023data}, who use global-local shrinkage priors to estimate cointegration relations in a data-driven manner.} As a solution, we adapt the approach of \citet{chakraborty2016bayesian} for the TVP-VECM and estimate ${r}_t$ for each period. 

It is convenient to consider a reparameterized version of Eq. (\ref{eq:tvpvecm}), where $\bm{\Pi}_t = \bm{\alpha}_t\bm{\beta}'$, see also \citet{liu1999parameter}. Here, the short-run adjustment coefficients are collected in $\bm{\alpha}_t$ of dimension $M\times q$, and the long-run relationships are captured by $\bm{\beta}$ which is $q\times q$. Note that we follow \citet{yang2018state} and assume the long-run relations to be constant over time.\footnote{Assuming both $\bm{\alpha}_t$ and $\bm{\beta}_t$ to vary over time further complicates achieving identification. First, note that $\bm{\Pi}_t=\bm{\alpha}_t\bm{\beta}_t' = \bm{\alpha}_t\bm{Q}\bm{Q}^{-1}\bm{\beta}_t'$ for any non-singular matrix $\bm{Q}$ which results in the so-called global identification problem. It is common in the literature to use linear normalization schemes such as $\bm{\beta}_t = (\bm{I}_{r_t},\bm{\beta}_t')'$, see also \citet{villani2001bayesian} or \citet{strachan2003valid}. Second, the local identification problem appears for $\bm{\alpha}_t=\bm{0}$ which implies that $\text{rank}(\bm{\Pi}_t) = 0$ \citep[see][]{kleibergen1994shape,kleibergen1998bayesian,paap2003bayes}.} This amounts to the assumption that long-run fundamental relations do not change over time, since our interest centers on the combined matrix $\bm{\Pi}_t$, where nonlinearities appear through $\bm{\alpha}_t$, which is sometimes also referred to as a loadings matrix. 

We refrain from restricting the cointegration space by imposing a deterministic structure on $\bm{\beta}$ \citep[see also][]{strachan2003valid,strachan2004bayesian,villani2006bayesian}. Instead, we follow \citet{koop2009efficient} and \citet{koop2011bayesian} and use the transformations ${\bm{\tilde\alpha}}_t = \bm{\alpha}_t\bm{\zeta}^{-1}$, $\bm{\tilde\beta} = \bm{\beta}\bm{\zeta}$, with $\bm{\zeta} = (\bm{\tilde\beta}'\bm{\tilde\beta})^{-0.5}$. This allows for employing a linear state-space modeling approach assuming conditional Gaussianity with a cointegration space prior. 

\subsection{Time-varying parameters and shrinkage}
Using $\bm{\tilde w}_t=\bm{\tilde\beta}'\bm{w}_t$ and $\bm{z}_t=(\bm{\tilde w}_t',\bm{x}_t')'$, a more compact version of Eq. (\ref{eq:tvpvecm}), providing notational simplicity, is given by:
\begin{equation}
\Delta\bm{y}_t = \bm{B}_t \bm{z}_t + \bm{L}_t\bm{\eta}_t, \quad \bm{\eta}_t\sim\mathcal{N}(\bm{0},\bm{H}_t),\label{eq:tvpvecmsimple}
\end{equation}
with $\bm{A}_t=(\bm{A}_{1t},\hdots,\bm{A}_{Pt},\bm{\gamma}_t)$, $\bm{B}_t=(\bm{\tilde\alpha}_t,\bm{A}_t)$ and $\bm{x}_t=(\Delta\bm{y}'_{t-1},\hdots,\Delta\bm{y}'_{t-P},\bm{c}_t')'$, where $\bm{A}_t$ is of size $M\times J$ and $\bm{x}_t$ of size $J\times1$ with $J=(MP+N)$. Furthermore, it is convenient to factor $\bm{\Sigma}_t = \bm{L}_t \bm{H}_t \bm{L}_t'$, with a diagonal matrix $\bm{H}_t =$ diag$(\exp(h_{1t}),\hdots,\exp(h_{Mt}))$ and $\bm{L}_t$ denoting the normalized lower Cholesky factor (i.e., a lower triangular matrix with ones on its diagonal). Note that Var$(\bm{L}_t\bm{\eta}_t) =$ Var$(\bm{\epsilon}_t)$; and this triangularization of the multivariate system allows for equation-by-equation estimation \citep[see, e.g.,][]{primiceri2005,carriero2019large, carriero2022corrigendum}.

We select the $i$th row of $\bm{B}_t$, and define $\bm{b}_{it} = \bm{B}_{i\bullet,t}'$ which refers to the parameters of the $i$th equation of the VECM. In addition, we stack all free elements of the matrix $\bm{L}_t$ in a vector $\bm{l}_t$. We then assume a random walk law of motion for these TVPs:
\begin{align}
\bm{b}_{it} &= \bm{b}_{it-1} + \bm{\vartheta}_{it}, \quad \bm{\vartheta}_{it}\sim\mathcal{N}(\bm{0},\bm{\Theta}_{(b)i}),\label{eq:ss-model_states}\\
\bm{l}_t &= \bm{l}_{t-1} + \bm{\vartheta}_{t}, \quad \bm{\vartheta}_{t}\sim\mathcal{N}(\bm{0},\bm{\Theta}_{(l)}).\nonumber
\end{align}
The state innovation variances, which govern the amount of time variation, are collected in diagonal covariance matrices $\bm{\Theta}_{(b)i}$ and $\bm{\Theta}_{(l)}$. To impose shrinkage, we use the non-centered parameterization of the TVP model \citep[see][for details]{fs_wagner}, which splits the parameters into a constant and time-varying part:
\begin{equation*}
\bm{b}_{it} = \bm{b}_{i0} + \sqrt{\bm{\Theta}}_{(b)i}\tilde{\bm{b}}_{it},\quad \tilde{\bm{b}}_{it} = \tilde{\bm{b}}_{it-1} + \tilde{\bm{\vartheta}}_{it}, \quad \tilde{\bm{\vartheta}}_{it}\sim\mathcal{N}(\bm{0},\bm{I}), \quad \tilde{\bm{b}}_{i0} = \bm{0},
\end{equation*}
for $i = 1,\hdots,M,$ with an analogously transformed state equation for the parameters $\bm{l}_t$. This allows to impose shrinkage on the constant part of the coefficients $\bm{b}_{i0}$, with $j$th element $b_{ij,0}$ and the amount of time variation determined by $\sqrt{\theta}_{(b)ij}$, the $j$th diagonal element of $\sqrt{\bm{\Theta}}_{(b)i}$. 

The sparsification methods proposed in this paper (see Section \ref{sec:sparsification}) may be combined with any desired setup from the class of global-local shrinkage priors, see \citet{cadonna2020triple} for a review. The respective parameters, in our case, the constant part of the coefficients and the square root of the state innovation variances, are assumed to follow a Gaussian distribution with zero mean, and a global variance parameter pushes all coefficients strongly towards zero. These global parameters are multiplied with local scalings, which allow to pull prior mass away from zero for specific parameters even in cases where the underlying parameter vector is very sparse. Both of these shrinkage factors are equipped with another prior hierarchy, and shrinkage properties arise from choices about these mixing distributions. From this class of priors, we choose the horseshoe of \citet{carvalho2010horseshoe} for its lack of prior tuning parameters and excellent shrinkage properties.

\subsection{Heteroskedastic errors}
To address possible heteroskedasticity, we use stochastic volatility (SV) models for the structural errors. The logarithms of the diagonal elements of $\bm{H}_t$ follow independent AR(1) processes,
\begin{equation}
h_{it} = \mu_{i} + \phi_i (h_{it-1} - \mu_{i}) + \varsigma_i\xi_{it},\quad \xi_{it}\sim\mathcal{N}(0,1),\label{eq:sv}
\end{equation} 
where $\mu_{i}$ is the unconditional mean, $\phi_i$ is the persistence parameter and $\varsigma_i$ is the error variance of the log-volatility process. Note that $\bm{\eta}_t = (\eta_{1t},\hdots,\eta_{Mt})'$ in Eq. (\ref{eq:tvpvecmsimple}) features Gaussian errors. We also consider specifications where we replace this Gaussian with a t-distribution,
\begin{equation*}
\eta_{it} \sim t_{\nu_i}(0,\exp(h_{it})),
\end{equation*}
which renders the model even more flexible. Note that as the degrees of freedom $\nu_i\rightarrow\infty$, we obtain a Gaussian as the limiting case. In our applied work, we place a prior on the degrees of freedom and estimate them alongside all other parameters.

This completes our baseline model specification. Additional details about priors and the sampling algorithm are provided in Appendix \ref{app:A}. It is worth noting that our prior choices are mostly standard and result in a fairly straightforward Markov chain Monte Carlo (MCMC) sampling algorithm which provides draws from the joint posterior of all model parameters.

\section{Dynamic sparsification}\label{sec:sparsification}
The model proposed above combined with continuous global-local priors yields posterior draws that are pushed towards approximate sparsity.\footnote{As highlighted by \cite{hahncarvalho2015dss}, the success of the two-step \textit{shrink-then-sparsify} approach depends on the shrinkage properties of the prior. They note that the horseshoe prior is well suited for such procedures, which is another reason why we illustrate our proposed framework with this specific choice.} We rely on \textit{ex post} sparsification of these draws for each point in time to obtain exact sparsity. In cointegration models, particularly with TVPs, it is beneficial to adjust loss functions for specific parts of the parameter space. This is due to subtle implications of these blocks of parameters for the overall model structure.

\subsection{Designing suitable loss functions}
To perform variable selection and obtain sparse coefficient matrices we postprocess $\bm \Pi_t$, $\bm A_t$ and $\bm \Sigma_t$ by minimizing three coefficient-type specific \textsc{Lasso} loss functions. In particular, we rely on methods proposed in \cite{friedman2008sparse}, \cite{hahncarvalho2015dss}, \cite{chakraborty2016bayesian}, \cite{bhattacharya2018signal}, and \cite{bashir2019post}, which have been successfully used in a range of multivariate and univariate macroeconomic and finance applications \citep[see][]{puelz2017variable, puelz2019portfolio, huber2020inducing, hauzenberger2020combining}. Given the properties of the reduced rank matrix $\bm \Pi_t$, we propose modifications when compared to sparsification of the autoregressive coefficients, $\bm A_t$, and the covariance matrix $\bm \Sigma_t$. 

As a general remark on notation, draws from the non-sparsified posteriors are indicated by a hat (e.g., $\hat{\bm{\Pi}}_t$) and sparse estimates are marked with an asterisk (e.g., $\bm{\Pi}^{\ast}_t$). The key difference between the two is that the non-sparsified posterior draws may have many entries close to but not exactly zero, while the sparsified estimate has exact zeros which are imposed using auxiliary loss functions. The general idea is that these loss functions are designed to reward model fit by minimizing a distance measure between the non-sparse and sparse solutions, while an additional tuning parameter penalizes non-zero parameters. Choosing this tuning parameter --- which, loosely speaking, governs the number of zeroes --- is generally not a trivial task. To avoid extensive pre-estimation tuning procedures, we use the signal adaptive variable selection (SAVS) estimator proposed in \citet{bhattacharya2018signal}, which selects an appropriate amount of sparsity automatically. 

In a TVP context, dynamic sparsification poses some additional challenges with respect to how sparsity is imposed with respect to $t$:
\begin{enumerate}[align=left]
\item \textit{Ex post} sparsification is commonly applied to point estimators, such as the posterior median or mean. We will deviate from this procedure and solve the respective optimization problem for each draw from the posterior distribution. The procedure of \cite{woody2019model} is closely related to this approach. They provide a theoretical motivation of conducting uncertainty quantification of sparse posterior estimates.
\item Our loss functions will be defined in terms of full-data matrices instead of $t$-specific covariates. The latter might be considered as the natural candidate when transforming a TVP model to its static representation \citep[for details, see][]{chan2009efficient, hhko2019}. Commonly, sparsification is applied in standard regression frameworks with constant coefficients, implying that all information over time is considered. Using $t$-by-$t$ draws would result in dependence on a single observation in time $t$.
\end{enumerate}
We illustrate these issues and our solutions in more detail below in the context of sparsifying the cointegration matrix. It is worth noting that these concerns apply to all three blocks of the parameters that we intend to dynamically sparsify. 

\subsection{Sparsifying the cointegration matrix}
Our basic approach to sparsifying the cointegration relationships follows \citet{chakraborty2016bayesian}. \textit{Ex post} sparsfying $\bm{\Pi}_t$ is of crucial importance to obtain an estimate for the rank. Using solely a global-local prior, some estimates in $\bm{\Pi}_t$ are pushed towards zero, but they are never exactly zero. In the case of the cointegration matrix, this implies that $\bm{\Pi}_t$ would always be of full rank (i.e., $r = M$ for all $t$). The main goal is thus to minimize the predictive loss between a non-sparsified draw $\hat{\bm{\Pi}}_t$ and a column-sparse solution $\bm{\Pi}_t^{\ast}$. 

Our interest centers on dynamic sparsification to obtain a sparse cointegration matrix at each point in time. The loss function is specified in terms of the full-data matrix $\bm{W}$, a $T\times q$ matrix with $\bm{w}'_t$ in the $t$th row:
\begin{equation}\label{eq:PIsps}
\bm \Pi_t^{\ast}{'} = \min_{\bm \Pi_t} \left(\lVert \bm W \hat{\bm \Pi}_t' - \bm{W}\bm{\Pi}_t' \rVert_F^{2}
+ \sum_{j = 1}^{q}  \kappa_{jt}  \lVert \bm{\Pi}_{\bullet j,t} \rVert_2 \right),
\end{equation}
with $\lVert \bm C \rVert_F$ denoting the Frobenius norm of a matrix $\bm C$, $\lVert \bm c \rVert_2$ the Euclidean norm of a vector $\bm c$ and $\bm \Pi_{\bullet j,t}$ referring to the $j$th column of $\bm \Pi_t$ (i.e., the $j$th row of its transpose). Equation \eqref{eq:PIsps} denotes a grouped \textsc{Lasso} problem with a row (column)-specific penalty $\kappa_{jt}$ and aims at finding a row (column)-sparse solution of $\bm{\Pi}'_t$ ($\bm{\Pi}_t$).\footnote{The Frobenius norm is a common distance measure between subspaces and is given by $\lVert \bm C \rVert_F = \sqrt{\text{tr}(\bm C' \bm C)}$ with $\text{tr}(\bm C)$ denoting the trace of a matrix $\bm C$. For a detailed discussion on properties of the grouped \textsc{Lasso}, see \cite{yuan2006model} and \cite{wang2008note}.} 

The first part controls the distance between an estimate and its sparse solution (measured by the Frobenius norm), while the second part penalizes non-zero elements in $\bm \Pi_t$ (in terms of column-specific Euclidean norms). We use the grouped \textsc{Lasso} as opposed to an element-wise \textsc{Lasso} to establish a loss function that penalizes the cointegration matrix towards a lower rank structure. It is worth noting that using an element-wise \textsc{Lasso} could yield situations where the penalty introduces spurious cointegration relationships. Summarizing in simple terms, this loss function is designed to obtain an adequate reduced rank estimate of $\bm{\Pi}_t$ that avoids sacrificing model fit relative to the full rank case.

Notice that we rely on the full data matrix $\bm W$ instead of the $t$-specific covariates $\bm w_t'$. This is in line with our state equations for the TVPs, where all information over time is used for filtering and smoothing. In constant parameter regressions, losses would be based on variation explained by $\bm W \bm \Pi'$. Using the static regression framework to perform dynamic sparsification and solely relying the $t$th observation $\bm w_t'$, instead of the full data matrix $\bm W$, would make the penalty highly sensitive to individual observations over time. To illustrate this, we focus on the $j$th columns in both $\bm W_{\bullet j}$ and $\bm \Pi_{\bullet j}$. The norm of $\bm W_{\bullet j}$ is defined by $\lVert \bm W_{\bullet j} \rVert_2 = \sqrt{\sum_{t=1}^T w_{tj}^2}$. When using $t$-by-$t$ estimates independently, the norm of the $t$th observation is given by $\lVert w_{tj} \rVert_2 = \sqrt{w_{tj}^2}$. A simple solution for this issue would be to down-weight the penalty in Eq. (\ref{eq:PIthresh}) by a factor $T$. However, although theoretically in line with the sparsification techniques proposed in \cite{hahncarvalho2015dss}, this has the disadvantage of being exposed to idiosyncrasies of the $t$th observation.\footnote{This poses the risk of unstable penalties in Eq. (\ref{eq:PIsps}), which may be particularly problematic with seasonal patterns in the data such as is the case for electricity prices.} Therefore, using $\bm W$ is arguably the most practicable solution, where each $t$-specific estimate is used for the entire sample. 

Note that Eq. (\ref{eq:PIsps}) can be interpreted as minimizing the expected loss \citep[see][]{hahncarvalho2015dss}. Setting $\hat{\bm \Pi}_t = \bm \Pi_t^{(s)}$ with $(s)$ indicating the $s$th MCMC draw (rather than a posterior point estimate), has the attractive feature of allowing for uncertainty quantification about the rank of $\bm \Pi_t$ \citep[see also][]{huber2020inducing,hauzenberger2020combining}. 

In fact, the resulting reduced rank cointegration matrix can be used to extract a model-based estimate of the number of cointegration relationships. An estimate of this time-varying rank $r_t$ is obtained using:
\begin{equation*}
r_t = \sum_{i=1}^M\mathbb{I}\left(\mathfrak{s}_{it}>\varphi\right),
\end{equation*}
with $\mathbb{I}(\bullet)$ denoting the usual indicator function and $\mathfrak{s}_{it}$ for $i=1,\hdots,M,$ are the singular values of $\bm W \bm{\Pi}_t^{\ast}{'}$. We follow \citet{chakraborty2016bayesian} and define the rank as the number of non-thresholded singular values, with $\varphi$ defined as the largest singular value of the residuals of the full data specification of \autoref{eq:tvpvecm}, which corresponds to the maximum noise level.

To avoid cross-validation for the penalty term \citep[as in][]{hahncarvalho2015dss}, we rely on the SAVS estimator proposed in \citet{bhattacharya2018signal} and set the penalty term to $\kappa_{jt} = 1/\lVert \bm{\hat\Pi}_{\bullet j,t} \rVert_2^2$. This yields the following soft threshold estimate:\footnote{To solve the optimization problem in \autoref{eq:PIsps}, the SAVS estimator can be interpreted as special case of the coordinate descent algorithm \citep{friedman2007pathwise} by relying on a single iteration to obtain a closed-form solution. \cite{bhattacharya2018signal} and \cite{hauzenberger2020combining} both provide evidence that the coordinate descent algorithm already converges after the first pass through.}
\begin{equation}\label{eq:PIthresh}
\bm{\Pi}^{\ast}_{\bullet j,t} =
\begin{cases} \bm 0_{M \times 1}, &\text{if} \quad \frac{\kappa_{jt}}{2 \lVert \bm{\hat\Pi}_{\bullet j,t} \rVert_2} \geq \lVert \bm W_{\bullet j} \rVert_2^2, \\[1em]
\left(1 - \frac{\kappa_{jt}}{2 \lVert \bm W_{\bullet j} \rVert_2^2 \lVert \bm{\hat\Pi}_{\bullet j,t} \rVert_2} \right) \bm{\hat\Pi}_{\bullet j,t}, &\text{otherwise}.
\end{cases}
\end{equation}
As shown by \cite{bhattacharya2018signal} this choice of $\kappa_{jt}$ has properties similar to the adaptive \textsc{Lasso} proposed by \cite{zou2006adaptive}. To extend our discussion about specifying the penalty in terms of full data matrices with respect to SAVS, it is worth noting that the only part that changes is how the norm of the data is calculated ($\lVert \bm W_{\bullet j}  \rVert_2^2$ instead of $w_{jt}^2$). In this case, the penalty specified in \cite{chakraborty2016bayesian}, $\kappa_{jt} = 1/(\lVert \hat{\bm \Pi}_{\bullet j,t} \rVert_2^2$), can be used (without correcting for $T$). Note that this is a practicable solution from an applied perspective, since the norm over the full data matrix is more robust than considering a single observation in period $t$.

\subsection{Sparsifying the autoregressive coefficients}
For sparsifying the time-varying autoregressive coefficients, we define a full-data matrix $\bm{X}$ of dimension $T\times J$ with $\bm{x}_t'$ in the $t$th row, an $MJ\times1$ vector $\bm{a}_t = \text{vec}(\bm{A}_t')$ and the corresponding $TM \times MJ$ regressor matrix $\tilde{\bm X} = (\tilde{\bm x}_1', \dots, \tilde{\bm x}_T')'$ with $\tilde{\bm x}_t = (\bm I_M \otimes \bm x_t')$. 

We assume a loss function of the form:
\begin{equation}\label{eq:Asps}
\bm{a}^{\ast}_t = \min_{\bm a_t} \left(\frac{1}{2}\lVert \tilde{\bm X} \hat{\bm a}_t - \tilde{\bm X} \bm a_t \rVert_2^{2}
+ \sum_{j = 1}^{J}  \delta_{jt} |a_{jt}| \right),
\end{equation}
with $a_{jt}$ denoting the $j$th element of $\bm{a}_t$ and $\delta_{jt}$ a covariate-specific penalty. It is worth noting that we minimize a standard \textsc{Lasso}-type predictive loss where the first part controls the distance between an estimate and a sparse solution and the second part penalizes non-zero elements in $\bm a_t$, different to the sparsification of the cointegration relationships. 

An optimal choice for the penalty is $\delta_{jt} = 1/(|\hat a_{jt}|^2)$. We again rely on the soft threshold estimate implied by SAVS to obtain a sparse draw of $\bm{a}_t$:
\begin{equation}\label{eq:Athresh}
a^{\ast}_{jt} =
\begin{cases} 0, &\text{if} \quad \frac{\delta_{jt}}{|\hat a_{jt}|} \geq \lVert \tilde{\bm X}_{\bullet j} \rVert_2^2, \\[1em]
\left(1 - \frac{\delta_{jt}}{\lVert \tilde{\bm X}_{\bullet j} \rVert_2^2 |\hat a_{jt}|} \right) \hat a_{jt}, &\text{otherwise}.
\end{cases}
\end{equation}
As shown by \cite{bhattacharya2018signal} this choice of $\delta_{jt}$ again has properties similar to the adaptive \textsc{Lasso} proposed by \cite{zou2006adaptive}. 

\subsection{Sparsifying the covariance matrix}
A sparse draw of the covariance matrix can be obtained by relying on methods proposed in \cite{friedman2007pathwise} and \citet{bashir2019post}. This involves using the precision rather than the covariance matrix, as the precision defines the conditional independence structure of the variables. It is worth noting that directly postprocessing estimates of the covariance matrix can result in a rather dense precision matrix and, hence, induce spurious contemporaneous relationships. Sparse estimates of the precision matrix $\bm{\Sigma}_t^{-1}$ are based on the graphical \textsc{Lasso} penalty:
\begin{equation}\label{eq:Sigsps}
\bm{\Sigma}_t^{-1 \ast} = \min_{\bm{\Sigma}_t^{-1}} \left(\text{tr}\left(\bm{\Sigma}_t^{-1} \bm{\hat\Sigma}_t \right) - \log\det \left(\bm{\Sigma}_t^{-1}\right) + \sum_{i \neq j} \lambda_{ij,t} |{\sigma}_{ij,t}^{-1}|\right),
\end{equation}
with $\text{tr}(\bm C)$ and $\log \det(\bm C)$ denoting the trace and the log-determinant of a square matrix $\bm C$, {$\lambda_{ij,t}$ an element-specific \textsc{Lasso} penalty and $\sigma_{ij,t}^{-1}$ the $j$th element in the $i$th row of $\bm{\Sigma}_t^{-1}$.} Similar to \autoref{eq:PIsps} and \autoref{eq:Asps} the parts $\text{tr}\left(\bm{\Sigma}_t^{-1} \bm{\hat\Sigma}_t \right) - \log\det \left(\bm{\Sigma}_t^{-1}\right)$ are measures of fit, while  the third term penalizes non-zero elements in the precision matrix. Here, it is worth noting that if $\sigma_{ij,t}^{-1\ast}$ is set to zero, the $i$th and the $j$th endogenous variable in the system do not feature a contemporaneous relationship. {Thus, postprocessing estimates of the precision matrix capture the notion of obtaining a truly sparse set of relationships between elements in $\Delta \bm y_t$.} 

Following \cite{bashir2019post}, the penalty in \autoref{eq:Sigsps} is chosen as $\lambda_{ij,t} = 1/|\sigma_{ij,t}^{-1}|^{0.5}$, which constitutes a semi-automatic procedure to circumvent cross-validation. Here, we refrain from showing the exact form of the soft threshold estimates for each element in $\bm \Sigma_t^{-1 \ast}$  and refer to \cite{friedman2008sparse} instead, who define a set of soft threshold problems, similar to \autoref{eq:Athresh}, to solve for a optimal solution for each element in $\bm \Sigma_t^{-1}$ (and thus $\bm \Sigma_t$). We use the coordinate descent algorithm provided by \cite{glasso} and only iterate once in line with the SAVS estimator.\footnote{Alternatively, one could also regularize the precision matrix by writing $\bm \Sigma^{-1}_t$ as $M$-dimensional set of nodewise regressions by using the triangularization decomposition outlined by \citet{meinshausen2006high}. \citet{friedman2008sparse} and \citet{banerjee2008model} show that this approach constitutes a special case of \autoref{eq:Sigsps} with $M$ independent \textsc{Lasso} problems.}

Estimating the model proposed in Section \ref{sec:econometrics} yields MCMC draws from the posterior distributions of all relevant parameters. These are subsequently postprocessed using the procedures described above in Section \ref{sec:sparsification}. The baseline framework may be applied to any dataset where one suspects the presence cointegration relationships. In the next section, we discuss further specification details in the context of applying this framework to a study of electricity prices.

\section{Modeling European electricity prices}\label{sec:application}
We use our proposed model in two different applications related to European electricity prices. First, we use daily data from several European electricity markets jointly. This serves to illustrate our approach in terms of detecting suitable cointegration relationships and nonlinearities in these relationships between interconnected energy markets in different European countries. Second, we produce forecasts of hourly electricity prices one-day-ahead. Here, we perform an extensive OOS forecast exercise, selecting German hourly electricity prices as the market of interest. We then evaluate our approach against a large set of competing models. This exercise serves to demonstrate that flexibly controlling for cointegration patterns in data (in this case, between hours during the day), in the absence of prior knowledge of such relationships, is beneficial for forecast accuracy.

\subsection{Dataset}
For the first application, we use daily prices (in levels, averaged over the hours of the day) to estimate our model jointly for nine different regional markets: Baltics (BALT), Denmark (DK), Finland (FI), France (FR), Germany (DE), Italy (IT), Norway (NO), Sweden (SE) and Switzerland (CH); i.e., $M=9$. The data are available for the period from January $1$st, $2017$ to December $31$st, $2019$ in EUR per megawatt-hour (MWh). We follow the literature and choose day-ahead prices determined on a specific day for delivery in a certain hour on the following day. 

Prices for BALT and the Nordic countries (DK, FI, NO, and SE) are obtained from \textit{Nord Pool}; the German, Swiss and French hourly auction prices are from the power spot market of the \textit{European Energy Exchange} (EEX); for the Italian prices, we use the single national prices (PUN) from the Italian system operator \textit{Gestore dei Mercati Energetici} (GEM). We preprocess the data for daylight saving time changes to exclude the $25$th hour in October and to interpolate the $24$th hour in March. As additional exogenous factors, we consider daily prices for coal and fuel and interpolate missing values for weekends and holidays. In particular, we use the closing settlement prices for coal (LMCYSPT) and one month forward ICE UK natural gas prices (NATBGAS) due to their importance for the dynamic evolution of electricity prices and potential cointegration relationships (i.e., $q = M+2 = 11$). The model also features deterministic seasonal terms (encoded in $\bm{c}_t$) based on the respective day of the week.

In our second application, the forecast comparison, we choose hourly day ahead prices (in levels) for Germany as our main country of interest, and focus on daylight hours ($8$ a.m. until $6$ p.m.) and an average of the night hours \citep[see also][]{raviv2015forecasting}. We use a hold-out period of approximately a year and a half ranging from July $3$rd, 2018 to December $31$st, 2019 (in total $550$ observations). We estimate TVP-VECMs with the individual hours per day being treated as dependent variables such that $M=12$. Our exercise is based on a pseudo OOS simulation using a rolling window of $T=400$ observations at a time.\footnote{{The main reason for the use of a rolling rather than expanding window is to limit the computational burden. A rolling window implies quicker parameter change over the holdout than an expanding one. However, the daily frequency provides enough observations for making more abrupt parameter change --- to possibly be captured by the TVPs --- likely also for the rolling window.}} We consider one-step ahead predictions, which implies that we forecast each individual hour for the following day. All models (both daily and hourly) feature $P=2$ lags.

\subsection{Nonlinearities and cointegration in European electricity prices}\label{sec:insample}
One major feature of the proposed approach is that it remains relatively agnostic on the precise form of the cointegration and parameter space spanned by short-run coefficients. We do not rule out complex patterns \textit{a priori}, but our flexible modeling approach is also capable of supporting a fairly parsimonious specification when the data suggests so.

In this sub-section, we examine the sparsification patterns and nonlinearities when daily electricity price data from multiple electricity markets are modeled jointly with a sparse TVP-VECM. We first assess whether these electricity prices are indeed cointegrated and whether these cointegration relationships change over time. We roughly gauge the nonlinear features of the cointegration space by assessing estimates for a time-varying cointegration rank. Second, we assess the sparsification patterns and nonlinearities of static and dynamic interdependencies across the European electricity market. To this end, we examine the sparsified estimates of both the short-run adjustment coefficients (dynamic interdependencies) and the contemporaneous relationships (static interdependencies). Third and finally, we investigate heteroskedastic data features and the role of large variance shocks.

\begin{figure}[t]
\centering
\includegraphics[scale=0.4]{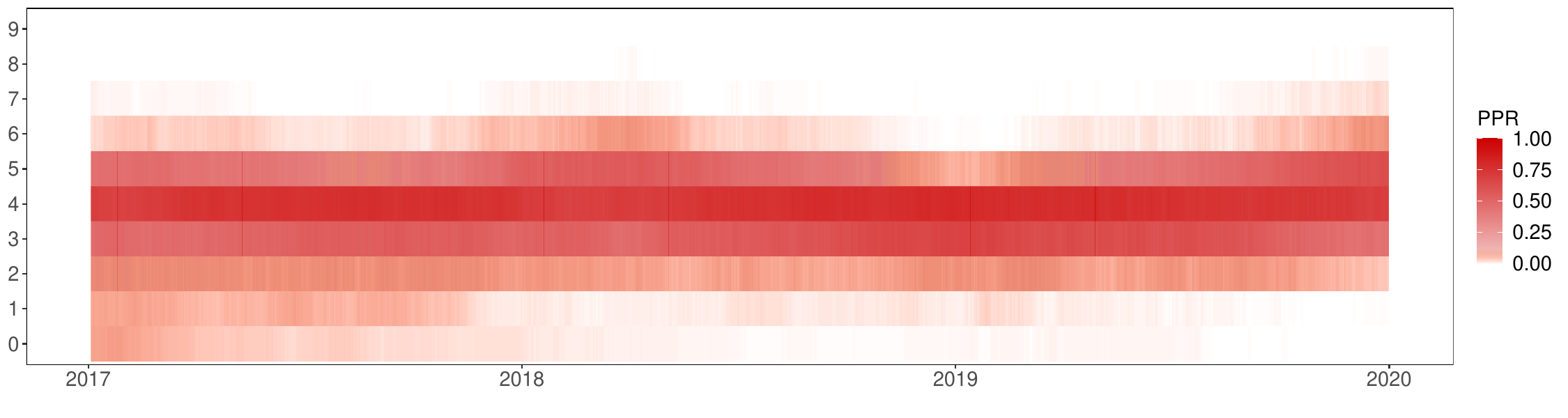}
\caption{Posterior probability of the rank (PPR) over time of the most flexible specification.}\label{fig:PPR}
\end{figure}

Figure \ref{fig:PPR} shows the posterior probability of the rank (PPR) over time based on our MCMC output. The probabilities are indicated in various shades of red. Several findings are worth noting. A rank {larger than six is hardly ever supported}, and while most posterior mass is concentrated on $r_t = 4$ for all $t$, we detect subtle differences over time.  At the beginning of the sample in $2017$, our estimates are more dispersed, with non-negligible probabilities for no cointegration. The precision of our rank estimate increases over time, with a much narrower corridor of probabilities starting around $2018$. {
After a brief period in late $2018$ and early $2019$ with probabilities shifting towards lower ranks, we find increases in cointegration relationships towards the end of the sample.
We conjecture that this pattern of the cointegration rank results from the fact that some of the countries (or regions) experience a large (idiosyncratic) positive shock in electricity prices in early 2018, while others do not. For example, electricity prices in the Baltics and in two Nordic countries (Finland and Sweden) increase from around 30 euros to 90 euros. In all other countries, these jumps are either less pronounced (e.g., Denmark and Norway) or almost nonexistent (e.g., Italy) during this episode.}

\begin{figure}[!ht]
\begin{minipage}{\textwidth}
\centering
(a) Coefficients linked to $\Delta \bm y_{t-p}$, for $p = 1, \dots, P$
\end{minipage}
\begin{minipage}{\textwidth}
\centering
\includegraphics[scale=0.30]{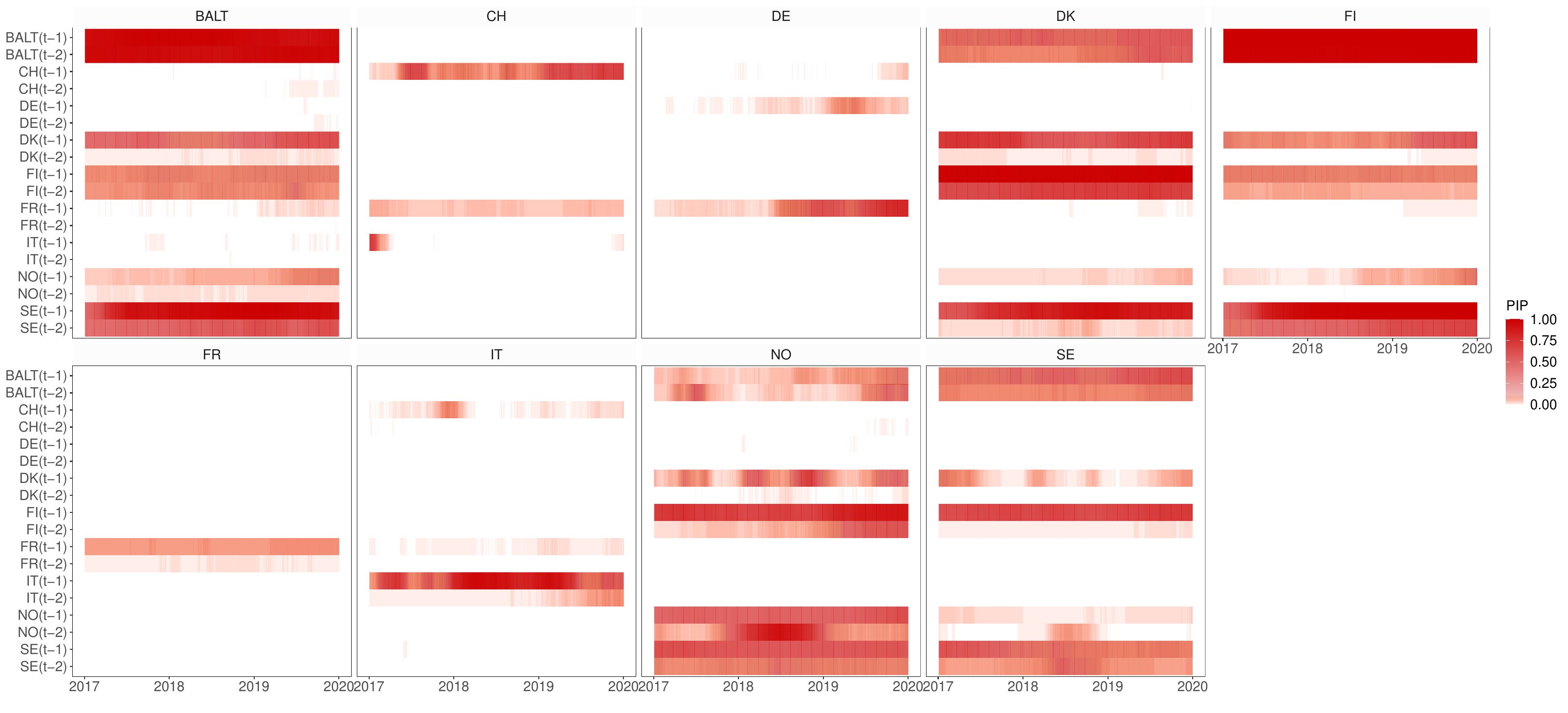}
\vspace*{1cm}
\end{minipage}
\begin{minipage}{\textwidth}
\centering
(b) Lower triangular part of variance covariance matrix 
\end{minipage}
\begin{minipage}{\textwidth}
\centering
\includegraphics[scale=0.36]{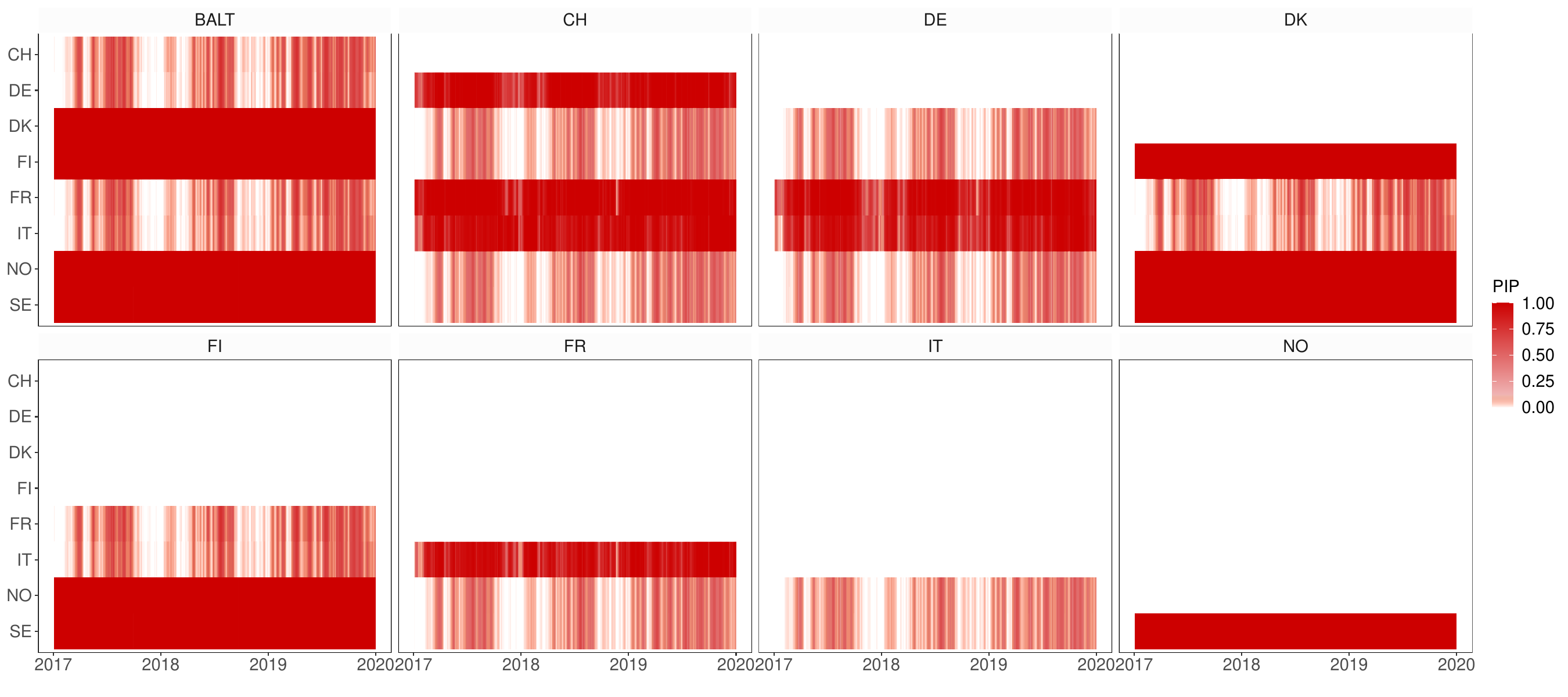}
\end{minipage}
\caption{Posterior inclusion probability (PIP) over time of autoregressive coefficients (panel (a)) and covariance matrices (panel (b)) of the most flexible specification.}\label{fig:coefficients}
\end{figure}

Next, we turn to sparsified estimates of the autoregressive coefficients and the error covariance matrix in panels (a) and (b) of Figure \ref{fig:coefficients}. Rather than showing magnitudes of the estimated coefficients, we use this exercise to illustrate the sparsification approach. As noted earlier, conventional shrinkage approaches push coefficients towards zero, but they are never exactly zero. The sparsification approach, on the other hand, introduces exact zeros in these matrices. We use this fact to compute the posterior inclusion probabilities (PIPs) of all coefficients by calculating the relative share of zeroes over the iterations of the algorithm. In other words, these figures showcase the sparsification approach as a variable selection tool \citep[see][]{hahncarvalho2015dss}.

We start with the coefficients linked to $\Delta\bm{y}_{t-p}$, that is, the dynamic interdependencies of the multivariate system in panel (a). A few findings are worth noting. First, we detect different degrees of sparsity across countries. While the Nordic and Baltic countries look rather similar (with comparatively dense coefficient matrices), this is not the case for Switzerland, Germany, Italy and France (CH, DE, FR, and IT). For these countries, the model estimates rather sparse coefficient matrices. Second, the variables with the highest PIPs are typically the countries' own lagged series. Particularly France shows extremely sparse estimates, with non-zero inclusion probabilities only for its own lags. Third, we observe several interesting changes in PIPs over time. This implies changes in the importance of predictors over time, a feature which our model detects in a data-based fashion. Fourth, we observe some noteworthy patterns of dynamic interdependencies, namely between Nordic and Baltic countries on one side, and to a lesser degree between continental European economies.

A similar exercise of PIPs in the context of the sparse covariance matrix is displayed in panel (b). Here, we show the lower triangular part of the contemporaneous relationships over time. As in the case of the regression coefficients, we detect differences in the degree of sparsity over the cross-section and across time. Strong contemporaneous relationships are detected especially between the Nordic countries. The PIPs in this case are often exactly one, indicating that the respective coefficients feature non-zero draws across all iterations of the sampler. Similarly, albeit with lower inclusion probabilities, we find that covariances appear to be important between continental European electricity prices. 

Interestingly, we observe a substantial degree of time-variation in the inclusion probabilities. Investigating these patterns in more detail, we find that the overall sparsity of the covariance matrix changes strongly over time, but does so in a specific way. In particular, there are periods where most covariance terms (apart from the always featured ones across Nordic countries) are sparsified strongly. Examples for such periods are in late $2017$ and early $2018$, or around the beginning of $2019$. Again, it is noteworthy that our model discovers these features in an automatic fashion.

We complete our discussion of the multi-country application by assessing whether heavy tailed errors are required to capture energy price fluctuations across the countries. For this purpose, we compare our estimates from our TVP-VECM benchmark model with SV to the same specification with t-distributed errors (see Appendix \ref{app:A}). We therefore assess the log-volatilities over time and across countries.

\begin{figure}[!ht]
\begin{minipage}{\textwidth}
\centering
(a) SV with t-distributed errors:
\end{minipage}
\begin{minipage}{\textwidth}
\centering
\includegraphics[scale=0.5]{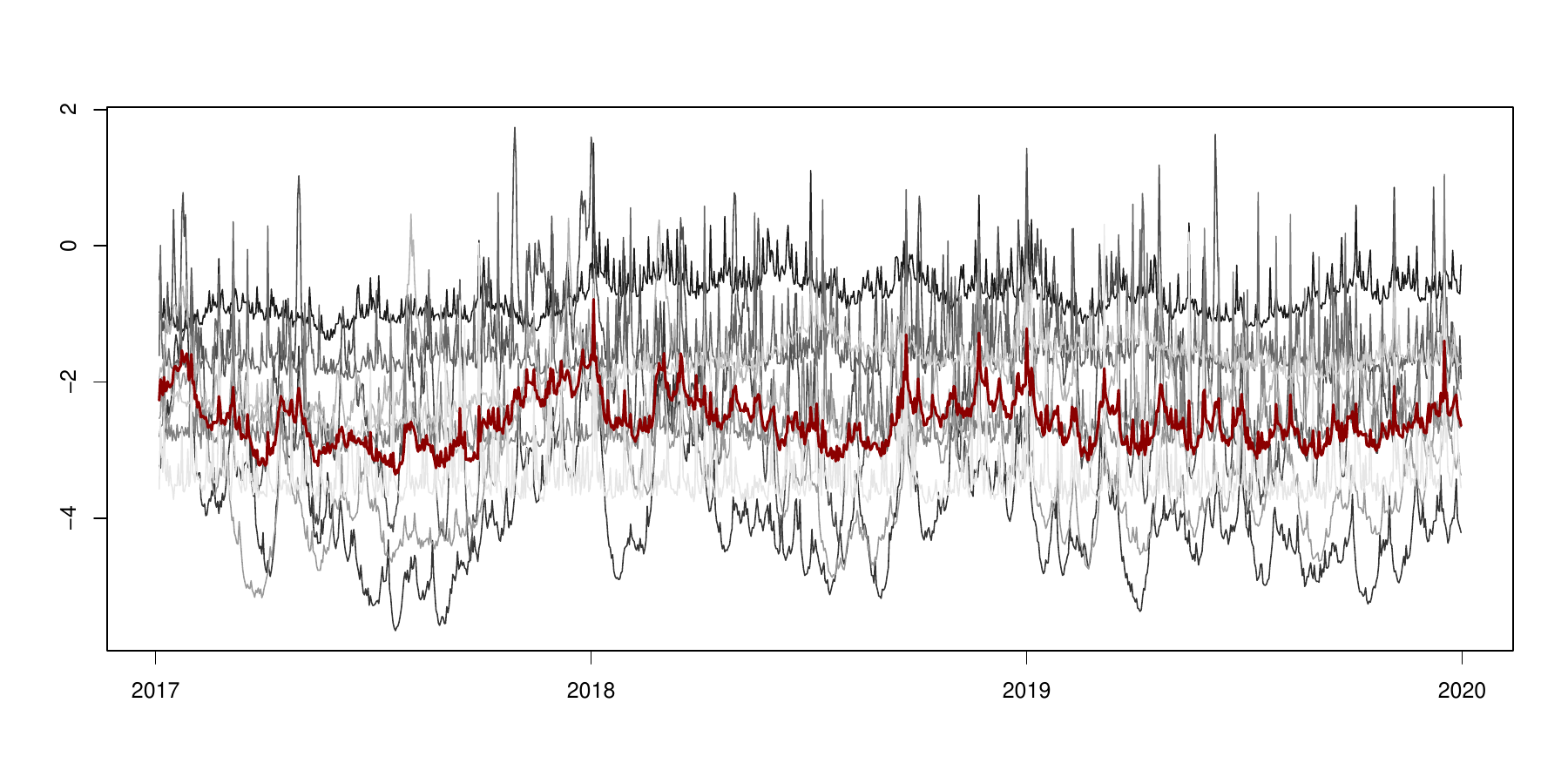}
\end{minipage}
\begin{minipage}{\textwidth}
\centering
(b) SV with Gaussian-distributed errors
\end{minipage}
\begin{minipage}{\textwidth}
\centering
\includegraphics[scale=0.5]{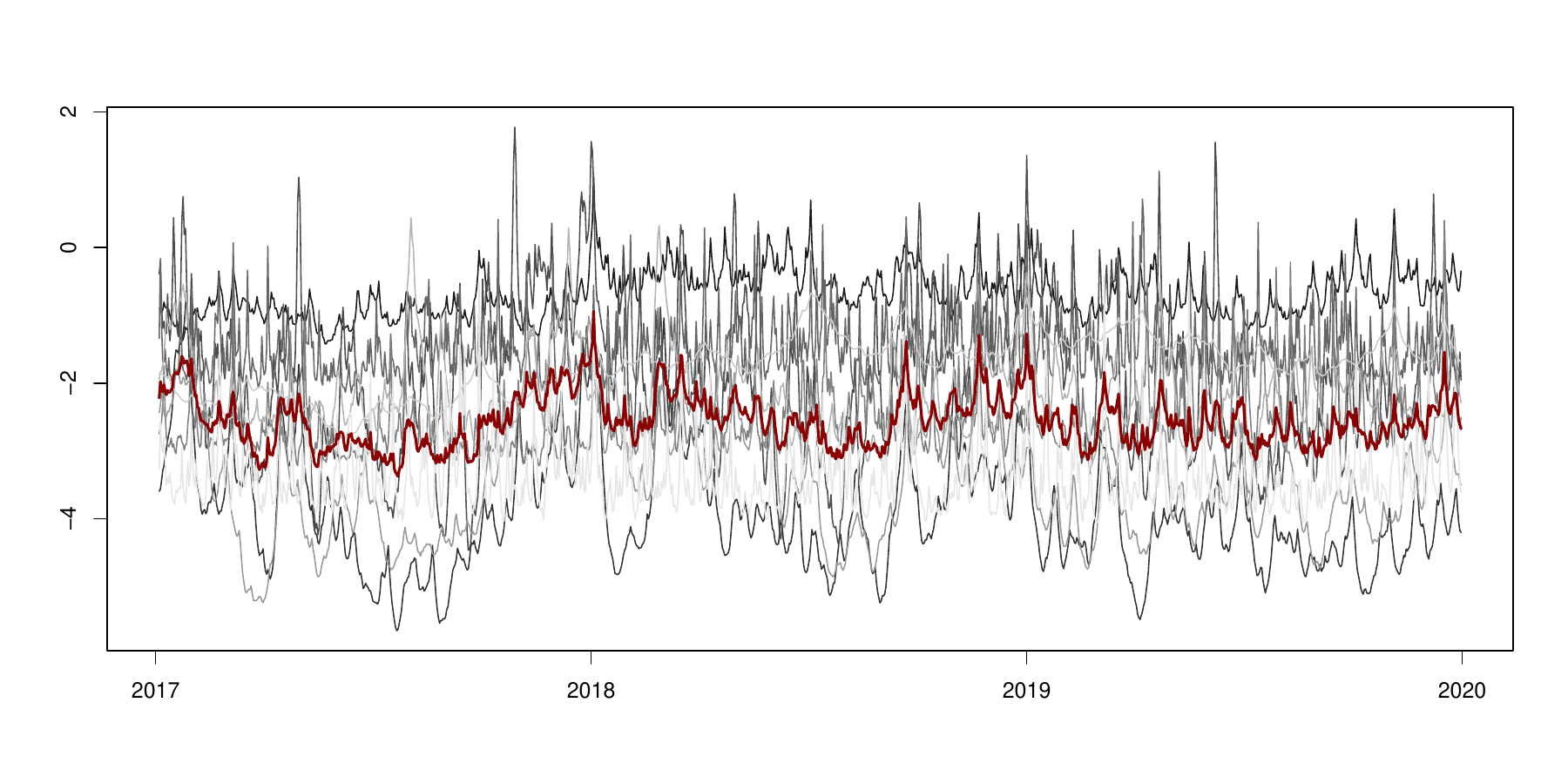}
\end{minipage}
\begin{minipage}{\textwidth}
\centering
\includegraphics[scale=0.7]{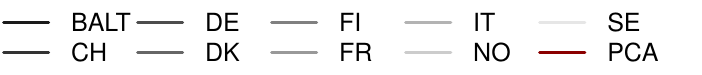}
\end{minipage}
\caption{Posterior median of SV. The red line denotes the first principal component (PCA) of $M$ latent processes ($h_{it}$).}\label{fig:svtime}
\end{figure}

Figure \ref{fig:svtime} shows our estimates for t-distributed errors in panel (a), while panel (b) indicates the standard SV specification. A few things are worth noting here. First, while the level of the volatilities varies substantially over the cross-section, the volatilities exhibit a substantial degree of co-movement. Second, even though we detect several differences between heavy tailed errors and conventional SVs, the first principal component of all volatility processes (marked by the red line) is almost identical for both error specifications. Third, t-distributed errors result in numerous high-frequency spikes for Denmark, Finland, and Sweden (i.e., the Nordic countries, apart from Norway). Table \ref{tab:svt-sv} in the Appendix displays the parameters of the state equation for the SVs and gives a more rigorous assessment of the dynamics in volatility processes. A crucial parameter in the context of heavy tailed errors is $\nu_{i}$, the degrees of freedom in the t-distribution. 

Overall, our results across countries are thus mixed. There is overwhelming empirical evidence for time-varying variances. On the other hand, we find that that t-distributed errors are not crucial for Baltic countries, Switzerland, Germany, France, Italy, and Norway, while the data supports heavier-than-normal tails for the remaining Nordic countries. Our model detects these features automatically, without the need of taking specific \textit{a priori} decisions about model specification. Next, we discuss whether these features in fact pay off in terms of predictive performance.

\subsection{Forecast results}\label{sec:forecasting}
In our forecasting comparison, we zoom in on a single country to conduct a thorough analysis featuring many competing models that have been shown to work well in the previous literature. In particular, we select hourly day ahead prices for Germany, and evaluate point and density forecasts for a set of models that are described in detail below.

\subsubsection{Competing models}
The competing specifications include a large set of univariate and multivariate models with constant parameters and TVPs. Moreover, we consider two variants for capturing heteroskedasticity. Our main interest centers on VECMs that are either sparsified or non-sparsified. For all these VECM specifications, we remain agnostic on the cointegration relations and estimate them from the data. As additional (non-sparsified) multivariate competitors, we consider VARs in levels and differences, while, as simple (non-sparsified) univariate competitors, we use AR($P$) models in levels and differences (with the latter model with constant coefficients serving as an overall benchmark). All these model variants are estimated with constant/time-invariant (\texttt{TIV}) and time-varying (\texttt{TVP}) parameters. With respect to the stochastic disturbances, we account for heteroskedasticity by relying either on a conventional SV specification with Gaussian errors (labeled \texttt{n} in the tables) as in Eq. (\ref{eq:sv}), or on an extension with t-distributed errors (labeled \texttt{t} in the tables) described in Appendix \ref{app:A}. Recall that all models feature two lags, i.e., two full days worth of hourly lags ($P=2$) and are equipped with a horseshoe prior \citep{carvalho2010horseshoe}. To economize on space, we refrain from showing results for homoskedastic specifications, since these models are typically found to be inferior for forecasting \citep[see, for instance,][]{gianfreda2020large}.

\subsubsection{Point and density forecasts}
Root mean squared errors (RMSEs) serve to evaluate the point predictions across our models. As a density forecast measure we rely on the continuous ranked probability score \citep[CRPS,][]{gneiting2007strictly}. This measure captures not only the first moment but also higher-order moments of the predictive distribution. Table \ref{tab:forecasts} displays RMSEs and CRPSs across all model types (rows) and over the respective hours of the day (columns). Forecasts are produced hourly for a full day ahead ($8$ a.m. until $6$ p.m., and aggregate ``Night'' hours). Besides individual hours, we produce a summary figure for overall forecast performance across a full day (labeled ``Total''). Model abbreviations and variants of features such as TVPs or the error variances specification are indicated in the previous sub-section. Values in the first row per model indicate RMSEs, those in parentheses in the second row are CRPSs. They are benchmarked relative (as ratios) to the AR($P$) model in differences with constant parameters and a standard SV specification (red shaded row, indicating raw values of RMSEs and CRPSs). In both cases, relative numbers below one mark superior forecast performance, with the best performing specification in bold.

The upper panel contains results for our main model variant (i.e., a VECM equipped with shrinkage priors) with both sparsified and non-sparsified estimates. The lower panel shows several non-sparsified benchmarks. It is worth mentioning that many of these benchmark specifications are nested in our proposed model. For example, in the case where our approach selects the cointegration matrix to be of full rank, we obtain a VAR in levels. Such specifications in essence differ in terms of the implied shrinkage prior on the reduced form coefficients \citep[see, for instance,][]{villani2009steady,eisenstat2016stochastic,giannone2019priors}.

Although our results indicate the lack of a single best performing specification, overall some interesting patterns emerge across the different model variants. While almost all multivariate models are superior compared to the univariate benchmarks, we obtain mixed evidence on whether one should choose VAR or VECM specifications to forecast electricity prices. Our results also indicate that taking into account cointegration relationships or directly modeling the data in levels is key to obtain accurate forecasts. Even a potentially too high number of cointegration relationships (by considering a VAR in levels) thus does not necessarily adversely affect predictive accuracy, whereas artificially removing all cointegration relationships (i.e., by considering a VAR in differences) results in a deteriorating forecast performance. By computing differences \textit{a priori}, i.e., naively transforming towards stationarity, rich information useful to improve forecasts may be lost.

Moreover, a flexible modeling approach for the conditional mean (i.e., by considering TVPs) often yields forecast gains. It is worth mentioning that this is also the case for comparatively worse performing model specifications, at least relative to the corresponding model with constant parameters. No clear picture emerges about the necessity of heavy tailed error distributions. This finding is in line with our discussion of insample findings for Germany in Section \ref{sec:insample}. Here, we found that when estimating a model with t-distributed errors, the large estimates for the degrees of freedom parameter suggests that a Gaussian error distribution is sufficient. It is worth mentioning, however, that t-distributed errors never substantially hurt predictive performance. 

While sparsification techniques do not necessarily improve point forecast accuracy, there is compelling evidence that these techniques are particularly beneficial for density forecast performance. This could be mainly explained by an underlying bias-variance trade-off, as the multiple loss functions involved in our \textit{shrink-then-sparsify} procedure strongly push coefficients of irrelevant predictors towards exact zeroes. Although this approach significantly reduces estimation uncertainty of the coefficients with the tighter predictive densities yielding improved density forecast performance, it may in turn introduce small biases resulting in a deterioration of point forecast performance.

Following this rather general discussion of our findings, we zoom into point and density forecast performance in more detail. Starting with point forecasts, Table \ref{tab:forecasts} indicates the non-sparsified TVP-VECM with Gaussian errors as the best performing specification. As suggested earlier, several other specifications, including our proposed model, display similar performance. These models show improvements upon the univariate benchmark model of about $15$ to $20$ percent lower RMSEs. This superior forecast performance is mainly driven by the excellent performance throughout the morning hours of the day. For night hours, on the contrary, improvements for point forecasts relative to the simple univariate benchmark are muted. We conjecture that electricity prices are more synchronized in the morning hours and late afternoon hours, where the demand for electricity (and thus prices) are typically high and thus more volatile, than during other hours of the day or at night. These considerations may provide an explanation why we observe the most pronounced gains in predictive accuracy with our proposed cointegrated approaches (or VAR models in levels) for (synchronised) peak hours, while the accuracy gains for (idiosyncratic) off-peak hours are rather muted. 

Turning to density forecasts in terms of CRPSs, we observe several differences compared to point forecast performance orderings. While most models show substantial improvements around $20$ percent lower CRPSs compared to the constant parameter AR($P$) model with conventional SVs, this metric selects our proposed model with Gaussian errors as the best performing specification for the full day (see column ``Total''). In particular, flexibly modeling the conditional mean and explicitly taking into account nonlinearities in the data pays off for density forecast performance. While allowing for nonlinear relationships in the conditional mean is beneficial, we observe that additional flexibility for the error variances (by considering t-distributed errors) is not necessarily needed.

\input{rmse_crps_new2.tex}

As is the case for point forecasts, we observe differences over the hours of the day. Our proposed sparsified TVP-VECMs show superior performance especially early in the morning and late in the afternoon. We conjecture that it is precisely at these hours that a single or a few cointgration relationships are the main driver of the synchronized price series. Since our sparsified TVP-VECM variant sufficiently accounts for nonlinear features and structural breaks in prices and, in addition, results in a relatively parsimonious and regularized cointegration space compared to a TVP-VAR in levels (i.e., the richest specification in terms of cointegration relationships) or non-sparsified TVP-VECMs, we suspect that this feature is the main reason for the excellent forecast performance. 

Finally, to gauge the statistical significance of our results, we conduct a more thorough analysis of the robustness of our density forecast metrics over the holdout. Here, we employ the model confidence set (MCS) procedure of \citet{hansen2011model}, with a loss function specified in terms of CRPSs (taking into account all the characteristics of predictive densities). Empty cells indicate that the corresponding model is eliminated from the MCS for the respective variable and loss function. The resulting MCS is displayed in Table \ref{tab:MCS}.

The MCS procedure yields a set of specifications which is a collection of models that contains the best ones with our pre-defined level of $75$ percent confidence. Given the informativeness of the data, this procedure may either select a single best performing specification, or a ranking of several comparable models on our chosen level of $25$ percent significance \citep[see][]{hansen2011model}. Considering the results of this exercise in Table \ref{tab:MCS}, we find that our proposed model performs well overall. The TVP-VECMs are always included in the MCS (i.e., among the statistically significant superior model set), and a single TVP-VECM variant appears among the top three ranked models for density forecasts in most hours. It is also worth mentioning that VECMs generally appear to yield superior forecasts to VARs. However, compared to our previous discussion of CRPSs, we observe some differences. The MCS procedure yields a different performance ranking in terms of significance when compared to evaluating solely CRPSs and producing a ranking in absolute terms. We conjecture that this is due to several periods in our holdout sample that affect the end-of-sample metrics shown in parentheses in Table \ref{tab:forecasts}, whereas the MCS procedure is more robust to such idiosyncrasies. In particular, for the sparsified TVP-VECMs we observe high ranks particularly during the afternoon in terms of density forecasts.

\input{rmse_crps_new_MCS_R025.tex}

\section{Concluding remarks}\label{sec:conclusion}
In this paper, we propose a TVP-VECM equipped with shrinkage priors and heteroskedastic errors. We then discuss methods for inducing sparsity. This framework is capable of introducing exact zeroes in cointegration relationships, autoregressive coefficients and the covariance matrix. The main idea is to start with a suitably flexible and sophisticated specification, and to impose data-driven sparsity on the parameter space to obtain the simplest adequate nested version. Moreover, our procedure yields estimates for a time-varying cointegration rank, without the need for introducing prior information on the cointegration relationships.

We estimate our model using daily and hourly day-ahead prices for different European electricity markets. In the empirical section, we illustrate some features of our approach insample using a multicountry dataset, and conduct an extensive OOS forecast comparison for Germany. Regarding the insample analysis, we detect several interesting time-varying patters of sparsity in the autoregressive coefficients and the covariance matrix. Our proposed detects such features of the data automatically. In addition, we find that our approach is competitive when forecasting hourly one-day ahead electricity prices compared to a large set of univariate and multivariate benchmarks.

\small{\setstretch{0.85}
\addcontentsline{toc}{section}{References}
\bibliographystyle{custom.bst}
\bibliography{tvpvecm}}\normalsize\clearpage

\begin{center}
\Large\sffamily\textbf{Online Appendix}  \\
\onehalfspacing\LARGE{\sffamily\textbf{\titlestring}}
\end{center}

\begin{appendices}\crefalias{section}{appsec}
\setcounter{equation}{0}
\renewcommand\theequation{A.\arabic{equation}}
\section{Econometric appendix}\label{app:A}
\subsection{Prior distributions}\label{subsec:horseshoe}
\begin{enumerate}[align=left]
\item The state innovation variances, which govern the amount of time variation, are collected in diagonal covariance matrix $\bm{\Theta}_{(b)i}$ with $j$th element $\theta_{(b)ij}$. On the constant part of the TVPs and these state innovation variances, we impose: 
\begin{align*}
b_{ij,0}&\sim\mathcal{N}(0,\psi_{(b)ij}^2\varrho_{(b)i}^2),\quad \psi_{(b)ij}\sim\mathcal{C}^{+}(0,1),\quad\varrho_{(b)i}\sim\mathcal{C}^{+}(0,1),\\
\sqrt{\theta}_{(b)ij}&\sim\mathcal{N}(0,\psi_{(\vartheta)ij}^2\varrho_{(\vartheta){i}}^2),\quad \psi_{(\vartheta)ij}\sim\mathcal{C}^{+}(0,1),\quad\varrho_{(\vartheta)i}\sim\mathcal{C}^{+}(0,1),
\end{align*}
with equation-specific global shrinkage parameters $\varrho_{(s)i}$ and local scalings $\psi_{(s)ij}$ with $s\in\{b,\vartheta\}$. The difference to otherwise identical horseshoe priors on the $\bm{l}_t$'s is that the global shrinkage parameters pool information about sparsity across all equations, i.e., there is no $i$ index in this case.
\item Following, \cite{koop2009efficient}, the prior on the transformed time-invariant cointegration vectors $\tilde{\bm \beta}$ is 
\begin{equation*}
\tilde{\bm \beta}_v = \text{vec}(\tilde{\bm \beta}) \sim \mathcal{N}(\bm 0, s_0 \bm I_{Mq}),
\end{equation*}
with $s_0 = 0.1$ being a weakly informative choice. 
Here it is worth noting, that similar to \cite{chakraborty2016bayesian} we refrain of using another shrinkage prior on elements in $\tilde{\bm \beta}$, due to the multiplicative and non-identified structure of $\bm \Pi_t = \tilde{\bm \alpha}_t \tilde{\bm \beta}'$. 

\item For the SV component we follow \cite{kastner2014ancillarity} and specify: $\mu_i\sim\mathcal{N}(0,100)$, $(\phi_i+1)/2\sim\mathcal{B}(5,1.5)$ and $\varsigma_i\sim\mathcal{N}(0,1)$, where $\mathcal{B}$ denotes the Beta distribution. The prior on the initial states is $h_{i0}\sim\mathcal{N}\left(\mu_i,\varsigma_i^2/(1-\phi_i)\right)$.
\end{enumerate}

\subsection{Sketch of the sampling algorithm}
\begin{enumerate}[align=left]
\item Writing the multivariate system in triangular form allows for updates of the covariances and conditional mean parameters equation-by-equation. Conditional on all other parameters and the TVPs, the constant part of the coefficients and the state innovation variances can be drawn from their conditionally conjugate Gaussian posterior distribution with well-known moments. 
\item The TVPs can subsequently be obtained by using a straightforward forward-filter backward-sampling (FFBS) algorithm conditional on all other parameters \citep[see][]{carterkohn,fs1994}.
\item These draws are used to update the global and local shrinkage factors for the horseshoe prior using the auxiliary representation of \citet{makalic2015simple}.
\item Transformed cointegration vectors conditional on all other parameters follow a multivariate Gaussian posterior distribution.
\item The SVs and the parameters of their state equations are sampled using the \texttt{R} package \texttt{stochvol} \citep{kastner2016dealing} based on the structural errors. This step involves drawing the degrees of freedom parameter for the t-distribution in the case of heavy tailed errors.
\item Sparsification is performed for each draw as outlined in detail in Section~\ref{sec:sparsification}. 
\end{enumerate}

\renewcommand\thetable{B.\arabic{table}}
\section{Additional empirical results}\label{app:B}
Results for the parameters in the state equation of the time-varying error variances are displayed in Table \ref{tab:svt-sv}.

A crucial parameter in the context of heavy tailed errors is $\nu_{i}$, the degrees of freedom in the t-distribution. Note that as $\nu_i \rightarrow \infty$, this distribution approaches a Gaussian. Consequently, large estimates imply that Gaussian errors are sufficient for capturing large swings in energy prices. For most countries, estimates center around $15$, resulting in an error distribution that looks almost like a Gaussian distribution, albeit with a bit heavier tails. This implies that information in our dataset suggests that t-distributed errors are no crucial feature for Baltic countries ($\nu = 13$), Switzerland ($\nu = 14$), Germany ($\nu = 16$), France ($\nu = 16$), Italy ($\nu = 17$) and Norway ($\nu = 15$). Differences appear for Denmark ($\nu = 7$), Finland ($\nu = 2$) and Sweden ($\nu = 4$), that is, the Nordic countries, apart from the case of Norway. Here, we observe degrees of freedom between $2$ and $7$, with a substantial number of observations in the tails of the error distribution.

Turning to estimates of the unconditional mean, persistence and errors of the state equation, we find that for the countries exhibiting less need for heavy tails indicated by high estimates of $\nu_i$, the posterior moments in Table \ref{tab:svt-sv}, are almost identical. This is due to the more flexible error distribution approaching Gaussianity. For the Nordic countries apart from Norway, we detect a pattern in terms of differences between parameter estimates. We observe substantially more persistent SV processes for the heavy tailed specification, with lower variances for Denmark, Finland and Sweden. This implies that the amount of clustering in volatilities is heterogeneous across markets. Differences in the unconditional mean are muted, implying that these findings apply mainly to the dynamic properties of the underlying volatilities rather than heterogeneity in overall levels.

\begin{landscape}\begin{table*}[ht]
\caption{Posterior estimates for the state equations of the error variances.}\label{tab:svt-sv}\vspace*{-1.5em}
\begin{footnotesize}
\begin{center}
\begin{threeparttable}
\begin{tabular*}{\linewidth}{@{\extracolsep{\fill}} rccccccccc}
  \toprule
 & \multicolumn{9}{c}{\textbf{t-distributed errors}}\\
 \cmidrule{2-10}
  & \itshape BALT & \itshape  CH & \itshape  DE & \itshape  DK & \itshape  FI & \itshape  FR & \itshape  IT & \itshape  NO & \itshape SE \\ 
 \midrule
$\mu_i$ & -0.584 & -3.971 & -1.782 & -1.581 & -3.688 & -3.496 & -2.139 & -1.687 & -3.635 \\ 
 & (-0.835,-0.342) & (-4.264,-3.7) & (-2.066,-1.52) & (-2.052,-1.19) & (-4.068,-3.298) & (-3.802,-3.218) & (-2.327,-1.974) & (-1.937,-1.435) & (-4.194,-3.263) \\ 
  $\phi_i$ & 0.963 & 0.908 & 0.827 & 0.654 & 0.92 & 0.919 & 0.862 & 0.972 & 0.629 \\ 
  & (0.93,0.987) & (0.866,0.942) & (0.766,0.883) & (0.362,0.873) & (0.759,0.977) & (0.885,0.947) & (0.809,0.91) & (0.937,0.992) & (0.401,0.783) \\ 
$\varsigma_i^2$ & 0.125 & 0.459 & 0.742 & 0.364 & 0.17 & 0.431 & 0.398 & 0.103 & 0.595 \\ 
& (0.068,0.189) & (0.364,0.574) & (0.609,0.875) & (0.042,0.718) & (0.1,0.309) & (0.346,0.535) & (0.313,0.474) & (0.05,0.183) & (0.322,0.806) \\ 
 $\nu_i$ & 13.022 & 13.691 & 15.745 & 6.872 & 2.211 & 16.218 & 17.008 & 14.497 & 4.417 \\ 
  & (8.316,18.307) & (7.015,19.346) & (9.796,19.612) & (2.822,15.421) & (2.012,2.574) & (11.11,19.69) & (12.837,19.849) & (8.966,19.589) & (2.276,7.675) \\ 
  \midrule
   & \multicolumn{9}{c}{\textbf{Gaussian errors}}\\
 \cmidrule{2-10}
  & \itshape BALT & \itshape  CH & \itshape  DE & \itshape  DK & \itshape  FI & \itshape  FR & \itshape  IT & \itshape  NO & \itshape SE \\ 
 \midrule
$\mu_i$ & -0.522 & -3.903 & -1.786 & -1.655 & -3.130 & -3.572 & -2.123 & -1.650 & -3.684 \\ 
& (-0.700,-0.330) & (-4.165,-3.627) & (-2.039,-1.534) & (-1.787,-1.525) & (-3.506,-2.771) & (-3.839,-3.291) & (-2.279,-1.952) & (-1.864,-1.416) & (-3.875,-3.407) \\ 
  $\phi_i$ & 0.924 & 0.876 & 0.807 & 0.238 & 0.346 & 0.893 & 0.826 & 0.938 & 0.309 \\ 
   & (0.825,0.981) & (0.816,0.924) & (0.749,0.864) & (0.073,0.382) & (0.211,0.466) & (0.843,0.932) & (0.765,0.882) & (0.796,0.988) & (0.184,0.428) \\ 
$\varsigma^2_i$ & 0.202 & 0.560 & 0.817 & 1.016 & 1.220 & 0.508 & 0.473 & 0.172 & 1.227 \\ 
& (0.092,0.351) & (0.436,0.720) & (0.682,0.961) & (0.905,1.143) & (1.006,1.489) & (0.404,0.633) & (0.381,0.578) & (0.07,0.391) & (1.047,1.386) \\ 
   \bottomrule
\end{tabular*}
\begin{tablenotes}[para,flushleft]
\scriptsize{\textit{Notes}: The parameter $\nu_i$ are the degrees of freedom for the t-distribution. If $\nu_i \rightarrow \infty$, the errors approach a Gaussian distribution. $\mu_i$ denotes the unconditional mean, $\phi_i$ the autoregressive parameter, and $\varsigma_i^2$  the state variances of the $M$ independent latent log-volatility processes $h_{it}$. Posterior median alongside $90\%$ credible sets in parentheses.}
\end{tablenotes}
\end{threeparttable}
\end{center}
\end{footnotesize}
\end{table*}\end{landscape}

\end{appendices}

\end{document}

%% file: rmse_crps_new2.tex
\begin{landscape}\begin{table*}[t]
\caption{Forecast performance for point and density forecasts in parentheses relative to the benchmark.}\vspace*{-1.5em}
\label{tab:forecasts}
\begin{tiny}
\begin{center}
\vspace*{10pt}
\begin{threeparttable}
\begin{tabular*}{\linewidth}{@{\extracolsep{\fill}} cclcccccccccccccc}
\toprule
\multicolumn{1}{l}{\bfseries }&\multicolumn{1}{c}{\bfseries Coefficients }&\multicolumn{1}{c}{\bfseries SV}&\multicolumn{1}{c}{\bfseries }&\multicolumn{13}{c}{\bfseries 1-day ahead}\tabularnewline
\cmidrule{5-17}
\multicolumn{1}{l}{}&\multicolumn{1}{c}{}&\multicolumn{1}{c}{}&\multicolumn{1}{c}{}&\multicolumn{1}{c}{Total}&\multicolumn{1}{c}{$8$ a.m.}&\multicolumn{1}{c}{$9$ a.m.}&\multicolumn{1}{c}{$10$ a.m.}&\multicolumn{1}{c}{$11$ a.m.}&\multicolumn{1}{c}{$12$ noon}&\multicolumn{1}{c}{$1$ p.m.}&\multicolumn{1}{c}{$2$ p.m.}&\multicolumn{1}{c}{$3$ p.m.}&\multicolumn{1}{c}{$4$ p.m.}&\multicolumn{1}{c}{$5$ p.m.}&\multicolumn{1}{c}{$6$ p.m.}&\multicolumn{1}{c}{Night}\tabularnewline
\midrule
\multicolumn{17}{c}{}\tabularnewline
\multicolumn{17}{c}{\bfseries Sparsified VECMs}\tabularnewline
\multicolumn{17}{c}{}\tabularnewline
   ~~&   TVP&   t&   &   0.83&   0.80&   0.79&   0.79&   0.79&   0.82&   0.83&   0.82&   0.82&   0.84&   0.86&   0.91&   0.91\tabularnewline
   ~~&   &   &   &   (0.82)&   (0.71)&   (0.72)&   (0.78)&   (0.79)&   (0.82)&   (0.85)&   (0.85)&   (0.84)&   (0.84)&   (0.85)&   (0.92)&   (0.94)\tabularnewline
\shadeRow   ~~&   TVP&   n&   &   0.82&   0.80&   0.78&   0.78&   0.78&   0.81&   0.83&   0.83&   0.82&   0.84&   0.86&   0.91&   0.91\tabularnewline
\shadeRow   ~~&   &   &   &   (\textbf{0.77})&   (\textbf{0.68})&   (\textbf{0.69})&   (0.73)&   (0.74)&   (0.78)&   (0.80)&   (0.79)&   (\textbf{0.78})&   (\textbf{0.80})&   (\textbf{0.82})&   (0.88)&   (0.87)\tabularnewline
   ~~&   TIV&   t&   &   0.82&   0.80&   0.78&   0.78&   0.78&   0.81&   0.83&   0.82&   0.82&   0.85&   0.87&   0.93&   0.91\tabularnewline
   ~~&   &   &   &   (0.82)&   (0.70)&   (0.70)&   (0.77)&   (0.80)&   (0.85)&   (0.86)&   (0.87)&   (0.86)&   (0.88)&   (0.88)&   (0.85)&   (0.87)\tabularnewline
\shadeRow   ~~&   TIV&   n&   &   0.81&   0.79&   0.77&   0.76&   0.76&   0.79&   0.82&   0.81&   0.81&   0.84&   0.85&   0.91&   0.89\tabularnewline
\shadeRow   ~~&   &   &   &   (0.81)&   (0.70)&   (0.71)&   (0.76)&   (0.79)&   (0.84)&   (0.85)&   (0.86)&   (0.85)&   (0.86)&   (0.87)&   (0.84)&   (\textbf{0.87})\tabularnewline
\multicolumn{17}{c}{}\tabularnewline
\multicolumn{17}{c}{\bfseries Non-sparsified VECMs}\tabularnewline
\multicolumn{17}{c}{}\tabularnewline
~~&   TVP&   t&   &   0.80&   0.76&   0.75&   0.75&   0.76&   0.79&   0.82&   \textbf{0.80}&   0.79&   0.82&   \textbf{0.84}&   0.90&   0.90\tabularnewline
~~&   &   &   &   (0.82)&   (0.71)&   (0.72)&   (0.78)&   (0.79)&   (0.83)&   (0.85)&   (0.85)&   (0.84)&   (0.84)&   (0.85)&   (0.91)&   (0.93)\tabularnewline
\shadeRow   ~~&   TVP&   n&   &   0.80&   0.76&   0.75&   0.75&   0.76&   0.79&   0.82&   0.81&   0.80&   0.83&   0.85&   0.90&   0.90\tabularnewline
\shadeRow   ~~&   &   &   &   (0.82)&   (0.75)&   (0.75)&   (0.78)&   (0.78)&   (0.82)&   (0.82)&   (0.83)&   (0.83)&   (0.84)&   (0.86)&   (0.90)&   (0.92)\tabularnewline
~~&   TIV&   t&   &   0.80&   0.75&   0.75&   0.75&   0.76&   0.79&   0.82&   0.80&   0.79&   0.83&   0.85&   0.91&   0.90\tabularnewline
~~&   &   &   &   (0.80)&   (0.70)&   (0.71)&   (0.76)&   (0.77)&   (0.81)&   (0.83)&   (0.82)&   (0.81)&   (0.83)&   (0.85)&   (0.89)&   (0.92)\tabularnewline
\shadeRow   ~~&   TIV&   n&   &   \textbf{0.79}&   0.75&   \textbf{0.74}&   \textbf{0.74}&   \textbf{0.75}&   \textbf{0.78}&   0.82&   0.80&   0.80&   0.82&   0.85&   0.90&   0.88\tabularnewline
\shadeRow   ~~&   &   &   &   (0.81)&   (0.74)&   (0.75)&   (0.76)&   (0.77)&   (0.80)&   (0.81)&   (0.82)&   (0.82)&   (0.84)&   (0.86)&   (0.89)&   (0.92)\tabularnewline
\midrule
\multicolumn{17}{c}{}\tabularnewline
\multicolumn{17}{c}{\bfseries Non-sparsified VARs, estimated in levels}\tabularnewline
\multicolumn{17}{c}{}\tabularnewline
~~&   TVP&   t&   &   0.81&   0.76&   0.76&   0.77&   0.79&   0.82&   0.83&   0.82&   0.81&   0.84&   0.87&   0.87&   0.90\tabularnewline
~~&   &   &   &   (0.82)&   (0.75)&   (0.76)&   (0.77)&   (0.78)&   (0.81)&   (0.81)&   (0.82)&   (0.82)&   (0.84)&   (0.87)&   (0.92)&   (0.93)\tabularnewline
\shadeRow   ~~&   TVP&   n&   &   0.82&   0.77&   0.77&   0.78&   0.80&   0.83&   0.84&   0.83&   0.82&   0.85&   0.88&   0.88&   0.90\tabularnewline
\shadeRow   ~~&   &   &   &   (0.78)&   (0.71)&   (0.72)&   (\textbf{0.73})&   (\textbf{0.73})&   (\textbf{0.77})&   (\textbf{0.77})&   (\textbf{0.78})&   (0.79)&   (0.81)&   (0.83)&   (0.87)&   (0.87)\tabularnewline
~~&   TIV&   t&   &   0.80&   \textbf{0.75}&   0.74&   0.76&   0.78&   0.81&   \textbf{0.81}&   0.81&   \textbf{0.79}&   \textbf{0.82}&   0.85&   \textbf{0.85}&   \textbf{0.87}\tabularnewline
~~&   &   &   &   (0.86)&   (0.72)&   (0.73)&   (0.80)&   (0.83)&   (0.89)&   (0.90)&   (0.91)&   (0.90)&   (0.92)&   (0.93)&   (0.90)&   (0.93)\tabularnewline
\shadeRow   ~~&   TIV&   n&   &   0.81&   0.76&   0.75&   0.77&   0.79&   0.82&   0.82&   0.82&   0.80&   0.83&   0.86&   0.85&   0.88\tabularnewline
\shadeRow   ~~&   &   &   &   (0.85)&   (0.73)&   (0.73)&   (0.80)&   (0.83)&   (0.88)&   (0.88)&   (0.89)&   (0.88)&   (0.91)&   (0.92)&   (0.90)&   (0.93)\tabularnewline
\multicolumn{17}{c}{}\tabularnewline
\multicolumn{17}{c}{\bfseries Non-sparsified VARs, estimated in first differences}\tabularnewline
\multicolumn{17}{c}{}\tabularnewline
   ~~&   TVP&   n&   &   0.96&   0.94&   0.94&   0.94&   0.95&   0.96&   0.96&   0.97&   0.96&   0.99&   1.00&   0.98&   0.99\tabularnewline
   ~~&   &   &   &   (0.99)&   (0.89)&   (0.89)&   (0.97)&   (0.98)&   (1.02)&   (1.01)&   (1.04)&   (1.03)&   (1.06)&   (1.05)&   (1.00)&   (1.00)\tabularnewline
\shadeRow   ~~&   TIV&   n&   &   0.96&   0.94&   0.94&   0.94&   0.95&   0.97&   0.97&   0.97&   0.96&   0.99&   1.00&   0.98&   1.00\tabularnewline
\shadeRow   ~~&   &   &   &   (1.04)&   (0.92)&   (0.93)&   (1.02)&   (1.03)&   (1.08)&   (1.07)&   (1.08)&   (1.07)&   (1.10)&   (1.09)&   (1.05)&   (1.06)\tabularnewline
\multicolumn{17}{c}{}\tabularnewline
\multicolumn{17}{c}{\bfseries Non-sparsified AR($P$) models, estimated in levels}\tabularnewline
\multicolumn{17}{c}{}\tabularnewline
~~&   TVP&   n&   &   0.86&   0.84&   0.82&   0.84&   0.86&   0.86&   0.86&   0.87&   0.86&   0.88&   0.90&   0.87&   0.92\tabularnewline
~~&   &   &   &   (0.82)&   (0.78)&   (0.78)&   (0.79)&   (0.81)&   (0.82)&   (0.82)&   (0.84)&   (0.83)&   (0.84)&   (0.85)&   (\textbf{0.83})&   (0.89)\tabularnewline
\shadeRow  ~~&   TIV&   n&   &   0.91&   0.91&   0.89&   0.91&   0.90&   0.89&   0.91&   0.90&   0.89&   0.92&   0.94&   0.92&   0.97\tabularnewline
 \shadeRow  ~~&   &   &   &   (0.91)&   (0.90)&   (0.88)&   (0.90)&   (0.89)&   (0.90)&   (0.91)&   (0.91)&   (0.90)&   (0.93)&   (0.94)&   (0.92)&   (1.00)\tabularnewline
\multicolumn{17}{c}{}\tabularnewline
\multicolumn{17}{c}{\bfseries Non-sparsified AR($P$) models, estimated in differences}\tabularnewline
\multicolumn{17}{c}{}\tabularnewline   
~~&   TVP&   n&   &   1.00&   1.00&   1.01&   1.00&   1.00&   1.00&   0.99&   1.00&   1.00&   1.01&   0.99&   0.99&   1.00\tabularnewline
~~&   &   &   &   (1.00)&   (1.00)&   (1.01)&   (1.00)&   (1.00)&   (1.00)&   (0.99)&   (1.00)&   (1.00)&   (1.00)&   (0.99)&   (0.99)&   (1.01)\tabularnewline
\shadeBench  ~~&   TIV&   n&   &   0.94&   1.09&   1.03&   0.99&   0.98&   0.94&   0.95&   0.96&   0.96&   0.90&   0.88&   0.85&   0.69\tabularnewline
\shadeBench  ~~&   &   &   &   (0.54)&   (0.68)&   (0.62)&   (0.58)&   (0.57)&   (0.54)&   (0.55)&   (0.54)&   (0.55)&   (0.52)&   (0.50)&   (0.49)&   (0.37)\tabularnewline
\bottomrule
\end{tabular*}
\begin{tablenotes}[para,flushleft]
\scriptsize{\textit{Notes}: Point forecasts are evaluated using root mean squared errors (RMSEs), density forecasts are continuous ranked probability scores (CRPSs). The red shaded rows denote the benchmark and displays actual values for RMSEs and CRPSs. All other models are shown as ratios to the benchmark. Relative numbers below one mark superior forecast performance, with the best performing specification in bold.}
\end{tablenotes}
\end{threeparttable}
\end{center}
\end{tiny}
\end{table*}\end{landscape}

%% file: rmse_crps_new_MCS_R025.tex
\begin{landscape}\begin{table*}[t]
\caption{Model confidence set (MCS) for density forecasts.}\vspace*{-1.5em}
\label{tab:MCS}
\begin{tiny}
\begin{center}
\vspace*{10pt}
\begin{threeparttable}
\begin{tabular*}{\linewidth}{@{\extracolsep{\fill}} cclcrrrrrrrrrrrrr}
\toprule
\multicolumn{1}{l}{\bfseries }&\multicolumn{1}{c}{\bfseries Coefficients }&\multicolumn{1}{c}{\bfseries SV}&\multicolumn{1}{c}{\bfseries }&\multicolumn{13}{c}{\bfseries 1-day ahead}\tabularnewline
\cmidrule{5-17}
\multicolumn{1}{l}{}&\multicolumn{1}{c}{}&\multicolumn{1}{c}{}&\multicolumn{1}{c}{}&\multicolumn{1}{c}{Total}&\multicolumn{1}{c}{$8$ a.m.}&\multicolumn{1}{c}{$9$ a.m.}&\multicolumn{1}{c}{$10$ a.m.}&\multicolumn{1}{c}{$11$ a.m.}&\multicolumn{1}{c}{$12$ noon}&\multicolumn{1}{c}{$1$ p.m.}&\multicolumn{1}{c}{$2$ p.m.}&\multicolumn{1}{c}{$3$ p.m.}&\multicolumn{1}{c}{$4$ p.m.}&\multicolumn{1}{c}{$5$ p.m.}&\multicolumn{1}{c}{$6$ p.m.}&\multicolumn{1}{c}{Night}\tabularnewline
\midrule
\multicolumn{17}{c}{}\tabularnewline
\multicolumn{17}{c}{\bfseries Sparsified VECMs}\tabularnewline
\multicolumn{17}{c}{}\tabularnewline
~~&   TVP&   t&   &   6&  12&  11&  7&  6&  6&  4&  6&  4&  4 &  7&  8&  8\tabularnewline
\shadeRow ~~&   TVP&   n&   &   4&  11&  12&  4& \textbf{2}& \textbf{3}&   \textbf{2}&   4&   5&   5&   8&   6&   6\tabularnewline
~~&   TIV&   t&    &   7&   10&   9&   5&   5&   5&   \textbf{3}&   5&   6&   6&   6&   7&   9\tabularnewline
\shadeRow   ~~&   TIV&   n&   &   2&   9&   10&   \textbf{1}&   \textbf{1}&   \textbf{1}&   \textbf{1}&   \textbf{1}&   \textbf{2}&   \textbf{2}&   \textbf{2}&   5&   \textbf{2}\tabularnewline
\multicolumn{17}{c}{}\tabularnewline
\multicolumn{17}{c}{\bfseries Non-sparsified VECMs}\tabularnewline
\multicolumn{17}{c}{}\tabularnewline
~~&   TVP &   t&   &   5&   \textbf{3}&   \textbf{3}&   6&   8&   7&   7&   7&   7&   7&   4&   11&   10\tabularnewline
\shadeRow   ~~&   TVP&   n&   &   3&   \textbf{2}&   \textbf{2}&   \textbf{3}&   \textbf{3}&   4&   5&   \textbf{3}&   \textbf{3}&   \textbf{3}&   \textbf{3}&   10&   5\tabularnewline
   ~~&   TIV &   t&   &   8&   4&   4&   8&   7&   8&   11&   8&   8&   8&   5&   13&   13\tabularnewline
\shadeRow   ~~&   TIV &  n &   &   1&   \textbf{1}&   \textbf{1}&   \textbf{2}&   4&   \textbf{2}&   6&   \textbf{2}&   \textbf{1}&   \textbf{1}&   \textbf{1}&   14&   4\tabularnewline
\midrule
\multicolumn{17}{c}{}\tabularnewline
\multicolumn{17}{c}{\bfseries Non-sparsified VARs, estimated in levels}\tabularnewline
\multicolumn{17}{c}{}\tabularnewline
   ~~&   TVP &   t&   &   10&   5&   6&   10&   11&   12&   12&   12&   12&   12&   12&   4&   11\tabularnewline
\shadeRow   ~~&   TVP&   n&   &   9&   6&   5&   9&   9&   11&   10&   10&   11&   11&   11&   \textbf{3}&   12\tabularnewline
   ~~&   TIV&   t&   &   14&   7&   7&   13&   14&   14&   14&   14&   14&   14&   14&   12&   \textbf{3}\tabularnewline
\shadeRow   ~~&   TIV&   n&   &   13&   8&   8&   12&   13&   13&   13&   13&   13&   13&   13&   \textbf{2}&   \textbf{1}\tabularnewline
\multicolumn{17}{c}{}\tabularnewline
\multicolumn{17}{c}{\bfseries Non-sparsified VARs, estimated in first differences}\tabularnewline
\multicolumn{17}{c}{}\tabularnewline
~~&   TVP&   n&   &   &   &   &   &   &   &   &   &   &   &   &   &   \tabularnewline
\shadeRow ~~&   TIV&   n&   &   &   &   &   &   &   &   &   &   &   &   &   &   \tabularnewline
\multicolumn{17}{c}{}\tabularnewline
\multicolumn{17}{c}{\bfseries Non-sparsified AR($P$) models, estimated in levels}\tabularnewline
\multicolumn{17}{c}{}\tabularnewline
~~&   TVP&   n&   &   12&   &   &   11&   12&   9&   9&   11&   9&   9&   9&   \textbf{1}&   7\tabularnewline
\shadeRow ~~&   TIV&   n&   &   11&   &   &   &   10&   10&   8&   9&   10&   10&   10&   9&   14\tabularnewline
\multicolumn{17}{c}{}\tabularnewline
\multicolumn{17}{c}{\bfseries Non-sparsified AR($P$) models, estimated in differences}\tabularnewline
\multicolumn{17}{c}{}\tabularnewline
~~&   TVP&   n&   &   &   &   &   &   &   &   &   &   &   &   &   &   17\tabularnewline
\shadeRow ~~&   TIV&   n&   &   &   &   &   &   &   &   &   &   &   &   &   &   15\tabularnewline
\bottomrule
\end{tabular*}
\begin{tablenotes}[para,flushleft]
\scriptsize{\textit{Notes}: Results for the model confidence set (MCS) procedure of \citet{hansen2011model} at a $25$ percent significance level. The loss function is specified in terms of continuous ranked probability scores (CRPSs) as a density forecast measure. Empty cells indicate that the model is \textbf{not} part of the MCS. The top three ranked models are marked in bold.}
\end{tablenotes}
\end{threeparttable}
\end{center}
\end{tiny}
\end{table*}\end{landscape}

%% file: HPR_OS.bbl
\begin{thebibliography}{66}
\newcommand{\enquote}[1]{``#1''}
\providecommand{\natexlab}[1]{#1}

\bibitem[{Banerjee \emph{et~al.}(2008)Banerjee, Ghaoui, and
  d'Aspremont}]{banerjee2008model}
\textsc{Banerjee O, Ghaoui LE, and d'Aspremont A} (2008), \enquote{Model
  selection through sparse maximum likelihood estimation for multivariate
  Gaussian or binary data,} \emph{Journal of Machine Learning Research}
  \textbf{9}(Mar), 485--516.

\bibitem[{Bashir \emph{et~al.}(2019)Bashir, Carvalho, Hahn, and
  Jones}]{bashir2019post}
\textsc{Bashir A, Carvalho CM, Hahn PR, and Jones MB} (2019),
  \enquote{Post-Processing Posteriors Over Precision Matrices to Produce Sparse
  Graph Estimates,} \emph{Bayesian Analysis} \textbf{14}(4), 1075--1090.

\bibitem[{Bello and Reneses(2013)}]{Bello2013}
\textsc{Bello A, and Reneses J} (2013), \enquote{Electricity Price Forecasting
  in the Spanish Market using Cointegration Techniques,} in \enquote{The 33rd
  Annual International Symposium on Forecasting (ISF 2013) Forecasting with Big
  Data,} .

\bibitem[{Bosco \emph{et~al.}(2010)Bosco, Parisio, Pelagatti, and
  Baldi}]{Bosco2010}
\textsc{Bosco B, Parisio L, Pelagatti M, and Baldi F} (2010), \enquote{Long-Run
  Relations in {E}uropean Electricity Prices,} \emph{Journal of Applied
  Econometrics} \textbf{25}, 805--832.

\bibitem[{Bunea \emph{et~al.}(2012)Bunea, She, and Wegkamp}]{bunea2012joint}
\textsc{Bunea F, She Y, and Wegkamp MH} (2012), \enquote{Joint variable and
  rank selection for parsimonious estimation of high-dimensional matrices,}
  \emph{The Annals of Statistics} \textbf{40}(5), 2359--2388.

\bibitem[{Cadonna \emph{et~al.}(2020)Cadonna, Fr{\"u}hwirth-Schnatter, and
  Knaus}]{cadonna2020triple}
\textsc{Cadonna A, Fr{\"u}hwirth-Schnatter S, and Knaus P} (2020),
  \enquote{Triple the gamma--A unifying shrinkage prior for variance and
  variable selection in sparse state space and TVP models,} \emph{Econometrics}
  \textbf{8}(2), 20.

\bibitem[{Carriero \emph{et~al.}(2022)Carriero, Chan, Clark, and
  Marcellino}]{carriero2022corrigendum}
\textsc{Carriero A, Chan J, Clark TE, and Marcellino M} (2022),
  \enquote{Corrigendum to ``Large Bayesian vector autoregressions with
  stochastic volatility and non-conjugate priors"[J. Econometrics 212 (1)(2019)
  137--154],} \emph{Journal of Econometrics} \textbf{227}(2), 506--512.

\bibitem[{Carriero \emph{et~al.}(2019)Carriero, Clark, and
  Marcellino}]{carriero2019large}
\textsc{Carriero A, Clark TE, and Marcellino M} (2019), \enquote{Large Bayesian
  vector autoregressions with stochastic volatility and non-conjugate priors,}
  \emph{Journal of Econometrics} \textbf{212}(1), 137--154.

\bibitem[{Carter and Kohn(1994)}]{carterkohn}
\textsc{Carter C, and Kohn R} (1994), \enquote{On Gibbs sampling for state
  space models,} \emph{Biometrika} \textbf{81}(3), 541--553.

\bibitem[{Carvalho \emph{et~al.}(2010)Carvalho, Polson, and
  Scott}]{carvalho2010horseshoe}
\textsc{Carvalho CM, Polson NG, and Scott JG} (2010), \enquote{The horseshoe
  estimator for sparse signals,} \emph{Biometrika} \textbf{97}(2), 465--480.

\bibitem[{Chakraborty \emph{et~al.}(2020)Chakraborty, Bhattacharya, and
  Mallick}]{chakraborty2016bayesian}
\textsc{Chakraborty A, Bhattacharya A, and Mallick BK} (2020),
  \enquote{Bayesian sparse multiple regression for simultaneous rank reduction
  and variable selection,} \emph{Biometrika} \textbf{107}(1), 205--221.

\bibitem[{Chan and Jeliazkov(2009)}]{chan2009efficient}
\textsc{Chan JC, and Jeliazkov I} (2009), \enquote{Efficient simulation and
  integrated likelihood estimation in state space models,} \emph{International
  Journal of Mathematical Modelling and Numerical Optimisation}
  \textbf{1}(1-2), 101--120.

\bibitem[{Chua and Tsiaplias(2018)}]{chua2018bayesian}
\textsc{Chua CL, and Tsiaplias S} (2018), \enquote{A Bayesian Approach to
  Modeling Time-Varying Cointegration and Cointegrating Rank,} \emph{Journal of
  Business \& Economic Statistics} \textbf{36}(2), 267--277.

\bibitem[{de~Marcos \emph{et~al.}(2016)de~Marcos, Reneses, and
  Bello}]{Marcos2016}
\textsc{de~Marcos RA, Reneses J, and Bello A} (2016), \enquote{Long-term
  Spanish electricity market price forecasting with cointegration and VEC
  models,} in \enquote{2016 International Conference on Probabilistic Methods
  Applied to Power Systems (PMAPS),} 1--7.

\bibitem[{De~Vany and Walls(1999)}]{DeVany1999}
\textsc{De~Vany A, and Walls W} (1999), \enquote{Cointegration analysis of spot
  electricity prices: insights on transmission efficiency in the western US,}
  \emph{Energy Economics} \textbf{21}(5), 435--448.

\bibitem[{Eisenstat \emph{et~al.}(2016)Eisenstat, Chan, and
  Strachan}]{eisenstat2016stochastic}
\textsc{Eisenstat E, Chan JC, and Strachan RW} (2016), \enquote{Stochastic
  model specification search for time-varying parameter VARs,}
  \emph{Econometric Reviews} \textbf{35}(8-10), 1638--1665.

\bibitem[{Feldkircher \emph{et~al.}(2022)Feldkircher, Huber, Koop, and
  Pfarrhofer}]{feldkircher2022approximate}
\textsc{Feldkircher M, Huber F, Koop G, and Pfarrhofer M} (2022),
  \enquote{Approximate Bayesian inference and forecasting in huge-dimensional
  multicountry VARs,} \emph{International Economic Review} \textbf{63}(4),
  1625--1658.

\bibitem[{Friedman \emph{et~al.}(2007)Friedman, Hastie, H{\"o}fling, Tibshirani
  \emph{et~al.}}]{friedman2007pathwise}
\textsc{Friedman J, Hastie T, H{\"o}fling H, Tibshirani R,} \emph{et~al.}
  (2007), \enquote{Pathwise coordinate optimization,} \emph{The Annals of
  Applied Statistics} \textbf{1}(2), 302--332.

\bibitem[{Friedman \emph{et~al.}(2008)Friedman, Hastie, and
  Tibshirani}]{friedman2008sparse}
\textsc{Friedman J, Hastie T, and Tibshirani R} (2008), \enquote{Sparse inverse
  covariance estimation with the graphical lasso,} \emph{Biostatistics}
  \textbf{9}(3), 432--441.

\bibitem[{Friedman \emph{et~al.}(2019)Friedman, Hastie, and
  Tibshirani}]{glasso}
---{}---{}--- (2019), \emph{glasso: Graphical Lasso: Estimation of Gaussian
  Graphical Models}, {R} package version 1.11.

\bibitem[{Fr\"{u}hwirth-Schnatter(1994)}]{fs1994}
\textsc{Fr\"{u}hwirth-Schnatter S} (1994), \enquote{Data augmentation and
  dynamic linear models,} \emph{Journal of Time Series Analysis}
  \textbf{15}(2), 183--202.

\bibitem[{Fr\"{u}hwirth-Schnatter and Wagner(2010)}]{fs_wagner}
\textsc{Fr\"{u}hwirth-Schnatter S, and Wagner H} (2010), \enquote{Stochastic
  model specification search for Gaussian and partial non-Gaussian state space
  models,} \emph{Journal of Econometrics} \textbf{154}(1), 85--100.

\bibitem[{Geweke(1996)}]{geweke1996bayesian}
\textsc{Geweke J} (1996), \enquote{Bayesian reduced rank regression in
  econometrics,} \emph{Journal of Econometrics} \textbf{75}(1), 121--146.

\bibitem[{Gianfreda \emph{et~al.}(2019)Gianfreda, Parisio, and
  Pelagatti}]{gianfreda2019Energy}
\textsc{Gianfreda A, Parisio L, and Pelagatti M} (2019), \enquote{The
  {RES}-Induced Switching Effect Across Fossil Fuels: An Analysis of Day-Ahead
  and Balancing Prices,} \emph{The Energy Journal} \textbf{40}.

\bibitem[{Gianfreda \emph{et~al.}(in-press)Gianfreda, Ravazzolo, and
  Rossini}]{gianfreda2020large}
\textsc{Gianfreda A, Ravazzolo F, and Rossini L} (in-press), \enquote{Large
  time-varying volatility models for electricity prices,} \emph{Oxford Bulletin
  of Economics and Statistics} \textbf{forthcoming}.

\bibitem[{Giannone \emph{et~al.}(2019)Giannone, Lenza, and
  Primiceri}]{giannone2019priors}
\textsc{Giannone D, Lenza M, and Primiceri GE} (2019), \enquote{Priors for the
  long run,} \emph{Journal of the American Statistical Association}
  \textbf{114}(526), 565--580.

\bibitem[{Gneiting and Raftery(2007)}]{gneiting2007strictly}
\textsc{Gneiting T, and Raftery AE} (2007), \enquote{Strictly proper scoring
  rules, prediction, and estimation,} \emph{Journal of the American statistical
  Association} \textbf{102}(477), 359--378.

\bibitem[{Hahn and Carvalho(2015)}]{hahncarvalho2015dss}
\textsc{Hahn PR, and Carvalho CM} (2015), \enquote{Decoupling Shrinkage and
  Selection in Bayesian Linear Models: A Posterior Summary Perspective,}
  \emph{Journal of the American Statistical Association} \textbf{110}(509),
  435--448.

\bibitem[{Hansen \emph{et~al.}(2011)Hansen, Lunde, and Nason}]{hansen2011model}
\textsc{Hansen PR, Lunde A, and Nason JM} (2011), \enquote{The model confidence
  set,} \emph{Econometrica} \textbf{79}(2), 453--497.

\bibitem[{Hauzenberger \emph{et~al.}(2020)Hauzenberger, Huber, and
  Koop}]{hauzenberger2020dynamic}
\textsc{Hauzenberger N, Huber F, and Koop G} (2020), \enquote{Dynamic Shrinkage
  Priors for Large Time-varying Parameter Regressions using Scalable Markov
  Chain Monte Carlo Methods,} \emph{arXiv} \textbf{2005.03906}.

\bibitem[{Hauzenberger \emph{et~al.}(2022)Hauzenberger, Huber, Koop, and
  Onorante}]{hhko2019}
\textsc{Hauzenberger N, Huber F, Koop G, and Onorante L} (2022), \enquote{Fast
  and flexible Bayesian inference in time-varying parameter regression models,}
  \emph{Journal of Business \& Economic Statistics} \textbf{40}(4), 1904--1918.

\bibitem[{Hauzenberger \emph{et~al.}(2021{\natexlab{a}})Hauzenberger, Huber,
  and Onorante}]{hauzenberger2020combining}
\textsc{Hauzenberger N, Huber F, and Onorante L} (2021{\natexlab{a}}),
  \enquote{Combining Shrinkage and Sparsity in Conjugate Vector Autoregressive
  Models,} \emph{Journal of Applied Econometrics} \textbf{36}(3), 304--327.

\bibitem[{Hauzenberger \emph{et~al.}(2021{\natexlab{b}})Hauzenberger, Huber,
  Pfarrhofer, and Z\"orner}]{huber2018stochastic}
\textsc{Hauzenberger N, Huber F, Pfarrhofer M, and Z\"orner TO}
  (2021{\natexlab{b}}), \enquote{Stochastic model specification in Markov
  switching vector error correction models,} \emph{Studies in Nonlinear
  Dynamics \& Econometrics} \textbf{25}(2).

\bibitem[{Houllier and {De Menezes}(2012)}]{Houllier2012}
\textsc{Houllier MA, and {De Menezes} LM} (2012), \enquote{A fractional
  cointegration analysis of {E}uropean electricity spot prices,} in
  \enquote{9th International Conference on the European Energy Market,} .

\bibitem[{Huber \emph{et~al.}(2021)Huber, Koop, and
  Onorante}]{huber2020inducing}
\textsc{Huber F, Koop G, and Onorante L} (2021), \enquote{Inducing sparsity and
  shrinkage in time-varying parameter models,} \emph{Journal of Business \&
  Economic Statistics} \textbf{39}(3), 669--683.

\bibitem[{Huber \emph{et~al.}(2020)Huber, Koop, and
  Pfarrhofer}]{huber2020bayesian}
\textsc{Huber F, Koop G, and Pfarrhofer M} (2020), \enquote{Bayesian inference
  in high-dimensional time-varying parameter models using integrated rotated
  Gaussian approximations,} \emph{arXiv} \textbf{2002.10274}.

\bibitem[{Huber and Z{\"o}rner(2019)}]{huber2019threshold}
\textsc{Huber F, and Z{\"o}rner TO} (2019), \enquote{Threshold cointegration in
  international exchange rates: a Bayesian approach,} \emph{International
  Journal of Forecasting} \textbf{35}(2), 458--473.

\bibitem[{Jochmann and Koop(2015)}]{jochmann2015regime}
\textsc{Jochmann M, and Koop G} (2015), \enquote{Regime-switching
  cointegration,} \emph{Studies in Nonlinear Dynamics \& Econometrics}
  \textbf{19}(1), 35--48.

\bibitem[{Jochmann \emph{et~al.}(2013)Jochmann, Koop, Leon-Gonzalez, and
  Strachan}]{jochmann2013stochastic}
\textsc{Jochmann M, Koop G, Leon-Gonzalez R, and Strachan RW} (2013),
  \enquote{Stochastic search variable selection in vector error correction
  models with an application to a model of the UK macroeconomy,} \emph{Journal
  of Applied Econometrics} \textbf{28}(1), 62--81.

\bibitem[{Kastner(2016)}]{kastner2016dealing}
\textsc{Kastner G} (2016), \enquote{Dealing with stochastic volatility in time
  series using the R package stochvol,} \emph{Journal of Statistical Software}
  \textbf{69}(5), 1--30.

\bibitem[{Kastner and Fr{\"u}hwirth-Schnatter(2014)}]{kastner2014ancillarity}
\textsc{Kastner G, and Fr{\"u}hwirth-Schnatter S} (2014),
  \enquote{Ancillarity-sufficiency interweaving strategy (ASIS) for boosting
  MCMC estimation of stochastic volatility models,} \emph{Computational
  Statistics \& Data Analysis} \textbf{76}, 408--423.

\bibitem[{Kleibergen and Van~Dijk(1994)}]{kleibergen1994shape}
\textsc{Kleibergen F, and Van~Dijk HK} (1994), \enquote{On the shape of the
  likelihood/posterior in cointegration models,} \emph{Econometric Theory}
  \textbf{10}(3-4), 514--551.

\bibitem[{Kleibergen and Van~Dijk(1998)}]{kleibergen1998bayesian}
---{}---{}--- (1998), \enquote{Bayesian simultaneous equations analysis using
  reduced rank structures,} \emph{Econometric Theory} \textbf{14}(6), 701--743.

\bibitem[{Koop \emph{et~al.}(2009)Koop, Le{\'o}n-Gonz{\'a}lez, and
  Strachan}]{koop2009efficient}
\textsc{Koop G, Le{\'o}n-Gonz{\'a}lez R, and Strachan RW} (2009),
  \enquote{Efficient posterior simulation for cointegrated models with priors
  on the cointegration space,} \emph{Econometric Reviews} \textbf{29}(2),
  224--242.

\bibitem[{Koop \emph{et~al.}(2011)Koop, Leon-Gonzalez, and
  Strachan}]{koop2011bayesian}
\textsc{Koop G, Leon-Gonzalez R, and Strachan RW} (2011), \enquote{Bayesian
  inference in a time varying cointegration model,} \emph{Journal of
  Econometrics} \textbf{165}(2), 210--220.

\bibitem[{Liu and Wu(1999)}]{liu1999parameter}
\textsc{Liu JS, and Wu YN} (1999), \enquote{Parameter expansion for data
  augmentation,} \emph{Journal of the American Statistical Association}
  \textbf{94}(448), 1264--1274.

\bibitem[{Makalic and Schmidt(2015)}]{makalic2015simple}
\textsc{Makalic E, and Schmidt DF} (2015), \enquote{A simple sampler for the
  horseshoe estimator,} \emph{IEEE Signal Processing Letters} \textbf{23}(1),
  179--182.

\bibitem[{Meinshausen and B{\"u}hlmann(2006)}]{meinshausen2006high}
\textsc{Meinshausen N, and B{\"u}hlmann P} (2006), \enquote{High-dimensional
  graphs and variable selection with the lasso,} \emph{The Annals of
  Statistics} \textbf{34}(3), 1436--1462.

\bibitem[{Paap and Van~Dijk(2003)}]{paap2003bayes}
\textsc{Paap R, and Van~Dijk HK} (2003), \enquote{Bayes estimates of Markov
  trends in possibly cointegrated series: An application to US consumption and
  income,} \emph{Journal of Business \& Economic Statistics} \textbf{21}(4),
  547--563.

\bibitem[{Primiceri(2005)}]{primiceri2005}
\textsc{Primiceri G} (2005), \enquote{Time varying structural autoregressions
  and monetary policy,} \emph{Oxford University Press} \textbf{72}(3),
  821--852.

\bibitem[{Pr{\"u}ser(2023)}]{pruser2023data}
\textsc{Pr{\"u}ser J} (2023), \enquote{Data-based priors for vector error
  correction models,} \emph{International Journal of Forecasting}
  \textbf{39}(1), 209--227.

\bibitem[{Puelz \emph{et~al.}(2020)Puelz, Hahn, and
  Carvalho}]{puelz2019portfolio}
\textsc{Puelz D, Hahn PR, and Carvalho CM} (2020), \enquote{Portfolio selection
  for individual passive investing,} \emph{Applied Stochastic Models in
  Business and Industry} \textbf{36}(1), 124--142.

\bibitem[{Puelz \emph{et~al.}(2017)Puelz, Hahn, Carvalho
  \emph{et~al.}}]{puelz2017variable}
\textsc{Puelz D, Hahn PR, Carvalho CM,} \emph{et~al.} (2017), \enquote{Variable
  selection in seemingly unrelated regressions with random predictors,}
  \emph{Bayesian Analysis} \textbf{12}(4), 969--989.

\bibitem[{Raviv \emph{et~al.}(2015)Raviv, Bouwman, and
  Van~Dijk}]{raviv2015forecasting}
\textsc{Raviv E, Bouwman KE, and Van~Dijk D} (2015), \enquote{Forecasting
  day-ahead electricity prices: Utilizing hourly prices,} \emph{Energy
  Economics} \textbf{50}, 227--239.

\bibitem[{Ray and Bhattacharya(2018)}]{bhattacharya2018signal}
\textsc{Ray P, and Bhattacharya A} (2018), \enquote{Signal Adaptive Variable
  Selector for the Horseshoe Prior,} \emph{arXiv} \textbf{1810.09004}.

\bibitem[{Strachan(2003)}]{strachan2003valid}
\textsc{Strachan RW} (2003), \enquote{Valid Bayesian estimation of the
  cointegrating error correction model,} \emph{Journal of Business \& Economic
  Statistics} \textbf{21}(1), 185--195.

\bibitem[{Strachan and Inder(2004)}]{strachan2004bayesian}
\textsc{Strachan RW, and Inder B} (2004), \enquote{Bayesian analysis of the
  error correction model,} \emph{Journal of Econometrics} \textbf{123}(2),
  307--325.

\bibitem[{Villani(2001)}]{villani2001bayesian}
\textsc{Villani M} (2001), \enquote{Bayesian prediction with cointegrated
  vector autoregressions,} \emph{International Journal of Forecasting}
  \textbf{17}(4), 585--605.

\bibitem[{Villani(2006)}]{villani2006bayesian}
---{}---{}--- (2006), \enquote{Bayesian point estimation of the cointegration
  space,} \emph{Journal of Econometrics} \textbf{134}(2), 645--664.

\bibitem[{Villani(2009)}]{villani2009steady}
---{}---{}--- (2009), \enquote{Steady-state priors for vector autoregressions,}
  \emph{Journal of Applied Econometrics} \textbf{24}(4), 630--650.

\bibitem[{Wang and Leng(2008)}]{wang2008note}
\textsc{Wang H, and Leng C} (2008), \enquote{A note on adaptive group {L}asso,}
  \emph{Computational Statistics \& Data Analysis} \textbf{52}(12), 5277--5286.

\bibitem[{Weron(2014)}]{WERON20141030}
\textsc{Weron R} (2014), \enquote{Electricity price forecasting: A review of
  the state-of-the-art with a look into the future,} \emph{International
  Journal of Forecasting} \textbf{30}(4), 1030 -- 1081.

\bibitem[{Woody \emph{et~al.}(2021)Woody, Carvalho, and
  Murray}]{woody2019model}
\textsc{Woody S, Carvalho CM, and Murray JS} (2021), \enquote{Model
  interpretation through lower-dimensional posterior summarization,}
  \emph{Journal of Computational and Graphical Statistics} \textbf{30}(1),
  144--161.

\bibitem[{Yang and Bauwens(2018)}]{yang2018state}
\textsc{Yang Y, and Bauwens L} (2018), \enquote{State-space models on the
  Stiefel manifold with a new approach to nonlinear filtering,}
  \emph{Econometrics} \textbf{6}(4), 48.

\bibitem[{Yuan and Lin(2006)}]{yuan2006model}
\textsc{Yuan M, and Lin Y} (2006), \enquote{Model selection and estimation in
  regression with grouped variables,} \emph{Journal of the Royal Statistical
  Society: Series B (Statistical Methodology)} \textbf{68}(1), 49--67.

\bibitem[{Zou(2006)}]{zou2006adaptive}
\textsc{Zou H} (2006), \enquote{The adaptive Lasso and its oracle properties,}
  \emph{Journal of the American Statistical Association} \textbf{101}(476),
  1418--1429.

\end{thebibliography}
